\documentclass[aps,prd,superscriptaddress,twocolumn]{revtex4-2}
\usepackage{blindtext}

\usepackage{mathtools,mathalfa,amssymb,amsmath,amsfonts,amsthm,bm}
\usepackage[T1]{fontenc}
\usepackage{tipa}
\usepackage{wasysym}
\usepackage{appendix}
\usepackage{calrsfs}
\usepackage{tikz}
\usepackage[font=small,
   justification=justified,
   format=plain]{caption}
\usepackage{aas_macros}
\usetikzlibrary {decorations.pathmorphing}
\usepackage{orcidlink}
\usepackage{tensor}

\usepackage{soul}

\usepackage{hyperref}
\hypersetup{
    colorlinks = true,
    citecolor  = blue,
    linkcolor = blue,
    urlcolor  = blue,
    linkbordercolor = white
}
            
\usepackage{framed}
\usepackage{color}
\usepackage[mathscr]{euscript}

\newcounter{definition}
\definecolor{wuppergreen}{RGB}{137, 186, 23}

\newcommand{\pullback}[1]{\hbox{\lower0.5ex\hbox{${}_{\leftarrow}$}}\kern-1.9ex{#1}}
\newcommand{\pullbacklong}[1]{\hbox{\lower0.85ex\hbox{${}_{\longleftarrow}$}}\kern-3.0ex{#1}}
\newcommand{\pullbackllong}[1]{\hbox{\lower0.85ex\hbox{${}_{\longleftarrow\!\!-\!\!-\!\!-\!\!-}$}}\kern-6.4ex{#1}}

\usepackage{graphicx}
\usepackage{subcaption}
\DeclareMathOperator{\csch}{csch}

\begin{document}
\title{The real-time Feynman path integral for step potentials}

\author{Job Feldbrugge
\orcidlink{0000-0003-2414-8707}}
\email{job.feldbrugge@ed.ac.uk}
\affiliation{Higgs Centre for Theoretical Physics, University of Edinburgh, James Clerk Maxwell Building, Edinburgh EH9 3FD, UK}

\author{Ue-Li Pen \orcidlink{0000-0003-2155-9578}}
\email{pen@asiaa.sinica.edu.tw}

\affiliation{Canadian Institute for Theoretical Astrophysics, University of Toronto, 60 St. George Street, Toronto, ON M5S 3H8, Canada}
\affiliation{Department of Physics, University of Toronto, 60 St. George Street, Toronto, ON M5S 1A7, Canada}
\affiliation{Dunlap Institute for Astronomy \& Astrophysics, University of Toronto, AB 120-50 St. George Street, Toronto, ON M5S 3H4, Canada}
\affiliation{Institute of Astronomy and Astrophysics, Academia Sinica, Astronomy-Mathematics Building, No. 1, Section 4, Roosevelt Road, Taipei 10617, Taiwan}
\affiliation{Perimeter Institute for Theoretical Physics, 31 Caroline St. North, Waterloo, ON, Canada N2L 2Y5}
\affiliation{Canadian Institute for Advanced Research, CIFAR program in Gravitation and Cosmology}

\begin{abstract}
    Complex (semi-)classical paths, or instantons, form an integral part of our understanding of quantum physics. Whereas real classical paths describe classically allowed transitions in the real-time Feynman path integral, classically forbidden evolution is captured by complex semi-classical paths or instantons. In this paper, we uncover the rich, intricate nature of complex semi-classical paths and interference in the Feynman propagator of a non-relativistic quantum particle in both a smooth Woods-Saxon and a discontinuous Heaviside step potential. We demonstrate that the complex semi-classical paths are connected to caustics and may cease to exist as naive solutions to the boundary value problem when the semi-classical path encounters singularities of the potential. We generalise complex semi-classical paths to equivalence classes. Using this generalisation, we track the contribution of complex semi-classical paths beyond these singularity crossings and identify the instanton responsible for the quantum reflection. Whereas most complex contributions to the path integral are small, we demonstrate that in some cases the contribution of the complex semi-classical path is unsuppressed and persists into the semi-classical limit. Finally, we develop methods to detect the presence of complex semi-classical paths from propagation amplitudes. The structure of complex semi-classical paths and methods developed here generalises to a large set of problems in real-time quantum physics.
\end{abstract}

\maketitle

%%%%%%%%%%%%%%%%%%%%%%%%%%%%%%%
\section{Introduction}\label{sec:introduction}
In the Feynman path integral, classical paths emerge through constructive interference in the semi-classical limit \cite{Feynman:1948, Feynman:1965}. The Feynman path integral for the transition amplitude $G(x_1,x_0;T)$ to propagate between $x_0$ and $x_1$ in time $T$ is dominated by paths close to the classical paths. However, real classical paths do not capture the full nature of the quantum system. Classically forbidden transitions -- such as quantum tunnelling -- are generally governed by complex solutions of the classical equations of motion, known as \textit{complex semi-classical paths} or \textit{instantons}. These complex semi-classical paths are integral to the exploration of many quantum phenomena in our Universe -- including studies in quantum mechanics, quantum field theory \cite{Coleman:1978}, nuclear scattering theory \cite{Balian:1974}, and approaches to quantum gravity \cite{Gibbons:1993, Feldbrugge:2017}. Moreover, complex semi-classical paths may well stand at the foundation of a rigorous definition of the Feynman path integral as a functional integral \cite{Feldbrugge:2023c}. 

Recent studies have unveiled the rich nature of complex semi-classical paths. Complex semi-classical paths emerge from real classical paths through caustics \cite{Schulman:1975, Schulman:2012, Feldbrugge:2023a} as described by catastrophe theory \cite{Saunders:1980, Feldbrugge:2023d}. Following Picard-Lefschetz theory, a semi-classical path is relevant if and only if the associated steepest ascent manifold intersects the original integration domain \cite{Witten:2010, Feldbrugge:2017}. A complex solution becomes relevant/irrelevant to the path integral through the Stokes phenomena. Moreover, in a recent paper, we demonstrate that for models whose classical equations of motion are not of \textit{Painlevé type}, relevant complex semi-classical paths may be obscured by \textit{singularity crossings} \cite{Feldbrugge:2023a, Feldbrugge:2023b}. A complex semi-classical path can encounter a singularity of the analytic continuation of the potential as we smoothly change the external parameters. After crossing such a singularity, the path still contributes to the path integral but no longer exists as a naive complex solution to the boundary value problem. Rather, the complex classical path needs to be analytically continued across a branch cut. In this paper, we study the significance of complex semi-classical paths in both the smooth Woods-Saxon step and the discontinuous Heaviside step potential. These step potentials serve as an ideal model to study the nature of real-time path integrals, as they allow us to evaluate the path integral and the corresponding classical theory simultaneously. 

Step potentials are among the first problems one encounters while studying quantum mechanics \cite{Griffiths:2018}. They serve as a fundamental model to particle scattering, the behaviour of a wavefunction at discontinuities, and quantum phenomena that diverge from classical expectations. They are integral to the study of abrupt junctions in semiconductors or heterostructures in material science and scanning tunnelling microscopy, where electron transport across potential steps is critical for device operation. In optics, the quantum mechanical step potential models the propagation of light through a refractive index discontinuity, linking quantum mechanics to classical wave theory. Birrel and Davis use the Woods-Saxon potential to analyse particle creation in a curved space-time \cite{Birrell:1984}. At the same time, step potentials enable the mathematical exploration of the real-time Feynman path integral. After the harmonic oscillator, the linear potential and the Dirac delta potential \cite{Gaveau:1986}, step potentials were among the first quantum systems for which the real-time Feynman path integral was worked out. Barut and Duru were the first to develop a representation of the real-time path integral for the Heaviside step using generators of a canonical transformation \cite{Barut:1988}. Crandall proposed a combinatorial approach \cite{Crandall:1993}, de Carvalho uses a path decomposition expansion \cite{Carvalho:1993}, while Chetouani et al.\ \cite{Chetouani:1993} and de Aguiar \cite{Aguiar:1993} approached the path integral for the Heaviside step using spectral representations. Around the same time, Grosche solved the path integral for a particle in a smooth step \cite{Grosche:1989} realised by a Woods-Saxon potential \cite{Woods:1954} by relating it to the propagator for a particle in the famous P\"oschl-Teller potential. Subsequently, Grosche recovered the Heaviside step propagator with the Woods–Saxon propagator by considering the limit of an infinitely steep step \cite{Grosche:1993}. More recently, Acila et al.\ studied the propagator with a perturbative expansion \cite{Acila:2006}. Yearsley revisits the step potential via the path decomposition expansion \cite{Yearsley:2008}. Though insightful, these studies of the real-time propagator of step potentials have remained largely formal. 

In this paper, we derive the spectral representation of the real-time Feynman path integral of a particle in these step potentials and, for the first time, numerically evaluate the resulting integrals. Next, we reconstruct the interference pattern in terms of classical paths. We identify the caustics and Stokes phenomena, and singularity crossings (where the naive complex solution to the boundary value problem ceases to exist). We propose to generalise complex semi-classical paths to equivalence classes of classical paths. When the complex semi-classical path exists, our formalism reduces to the traditional picture. After the singularity crossing, the equivalence class enables the analytic continuation of the contribution of the complex semi-classical paths to the propagator. Interestingly, A. Kent already suggested that path integrals may receive contributions beyond the naive saddle points \cite{Kent:2013}. Note that, in classical mechanics, the equation of motion and the corresponding classical paths are fundamental. The action principle merely serves as an elegant method to derive the equations of motion through the Euler-Lagrange equation. In quantum physics, the roles are reversed. While the naive classical path loses its original interpretation when it becomes complex-valued and becomes even ill-defined at a singularity crossing, the associated action continues smoothly through the associated branch cut, remains well-defined and contributes to the propagator. Moreover, we show that in some circumstances, the contribution of the complex semi-classical path is unsuppressed (has a real classical action) and persists in the semi-classical limit. 

In this paper, we, for the first time, track the significance of a complex classical path directly from the propagator, using the Fourier transform and Laplace transform of the propagator with respect to the reciprocal of the reduced Planck constant $\omega=1/\hbar$. This approach complements previous proposals for the detection of relevant complex semi-classical paths based on weak value theory \cite{Aharonov:2003, Turok:2014, Feldbrugge:2019sew, Nishimura:2023} and may lead to the direct identification of real instantons from experiments. By uncovering the nature of complex semi-classical paths for the step potential, we hope to unlock their use in fields ranging from laser physics to quantum field theory and quantum gravity.

\subsection*{Outline}
In section \ref{sec:step_potential}, we present a new derivation of the real-time Feynman propagator for a particle in a smooth step realised by the Woods-Saxon potential. The presented method enables the evaluation of the spectral representation of the real-time path integral for any quantum mechanical model for which the time-independent Schrödinger equation has a closed-form solution. Though this method has likely been developed before, we are unaware of an analogous exposition. In section \ref{sec:Heaviside_step}, we take the limit of an infinitely steep step and recover the spectral representation of the Heaviside step propagator. The resulting propagator agrees with (most of the) previously reported propagators. The proposed spectral representations match the numerical evaluation of the real-time path integral \cite{Feldbrugge:2025}. Although the obtained spectral representation of these step propagators is not new, we include their derivation as the propagator of the Woods-Saxon potential was originally derived from the P\"oschl-Teller propagator instead of the spectral representation and the integral representations of the propagators play a crucial role in the subsequent analysis. For theories that do not allow for a closed-form spectral representation, we plan to use the recently developed numerical methods for discretised real-time path integrals \cite{Feldbrugge:2025}. Next, in section \ref{sec:classical-paths}, we evaluate the real-time propagator and investigate the resulting interference pattern in terms of the real classical paths and complex semi-classical paths solving the analogous boundary value problem. In particular, we generalise the complex solutions to the boundary value problem to an equivalence class of paths to traverse the singularity crossings. Finally, in section \ref{sec:fourier}, we study the relevance of classical solutions to the boundary value problem through the Fourier and Laplace transforms of the propagator with respect to the reciprocal of the Planck constant $\omega = 1/\hbar$. The analysis reaffirms the importance of complex semi-classical paths beyond singularity crossings. In section \ref{sec:applications}, we discuss the applications of the obtained results to and beyond the step potential and discuss future extensions of the presented exploration into quantum mechanics. We summarise our results in section \ref{sec:conclusion}.

%%%%%%%%%%%%%%%%%%%%%%%%%%%%%%%
\section{The smooth step}\label{sec:step_potential}
The Feynman path integral of a non-relativistic particle of mass $m$ in the potential $V$ expresses the propagator in terms of a sum over histories
\begin{align}
    G(x_1,x_0;T) = \int_{x(0)=x_0}^{x(T)=x_1} e^{i S[x]/\hbar}\mathcal{D}x\,,
\end{align}
interpolating between the initial $x_0$ and final position $x_1$, with the action
\begin{align}
    S[x] = \int_0^T \left(\frac{1}{2}m\dot{x}^2 - V(x)\right)\mathrm{d}t\,,
\end{align}
and the reduced Planck constant $\hbar$. The propagator is the Green's function of the Schrödinger equation, which satisfies the differential equation 
\begin{align}
    \left[i \hbar \frac{\partial}{\partial T} - \hat{H}\right] G(x_1,x_0;T) = i \hbar \delta(T)\,,
\end{align}
acting on either $x_0$ or $x_1$ (yielding two conditions), with the Hamiltonian operator
\begin{align}
    \hat{H} = -\frac{\hbar^2}{2m}\frac{\partial^2}{\partial x^2} + V(x)\,,
\end{align}
and the coincidence limit, 
\begin{align}
    \lim_{T\to 0} G(x_1,x_0;T) = \delta(x_1-x_0)\,,
\end{align}
with the Dirac delta function $\delta$. The convolution of an initial state $\psi_0$ (at time $t=0$) with the propagator 
\begin{align}
    \psi_t(x_1) = \int G(x_1,x_0;t)\psi_t(x_0)\mathrm{d}x_0\,,
\end{align}
yields the solution of the time-dependent Schrödinger equation 
\begin{align}
    i \hbar \frac{\partial \psi_t}{\partial t} = \hat{H} \psi_t\,,
\end{align}
with the initial condition $\psi_{t=0}(x)=\psi_0(x)$. The real-time path integral embodies quantum dynamics as it simultaneously captures all dynamical information of the quantum system in a single expression.

\begin{figure}
    \centering
    \includegraphics[width=\linewidth]{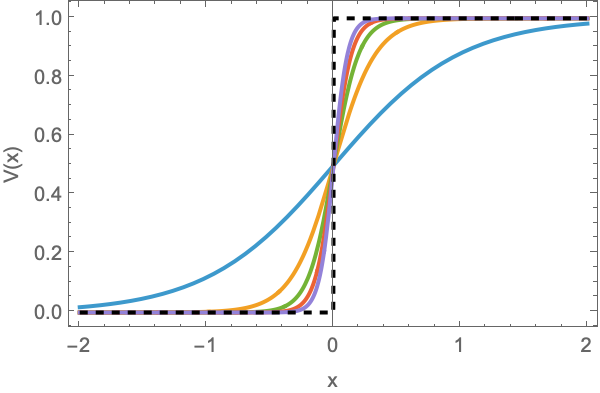}
    \caption{The smooth step potential with the strength $V_0=1$ and the steepness $\alpha=1, 3, \dots, 9$ (blue, yellow, green, red, and purple). The corresponding Heaviside step is plotted by the dashed black curve.}\label{fig:step}
\end{figure}

In this paper, we evaluate the real-time propagator for a particle encountering step potentials. We consider a particle in a Woods-Saxon \cite{Woods:1954} potential
\begin{align}
    V_{WS}(x)
    % &=\frac{V_0}{2}\left(1+\tanh \alpha x\right)\\
    &=\frac{V_0 }{1+e^{-2 \alpha x}}\,,
\end{align}
with the height of the step $V_0 >0$ and the steepness of the smooth step controlled by the parameter $\alpha >0$, and the famous Heaviside step potential 
\begin{align}
    V_H(x) = V_0 \Theta(x)\,,
\end{align}
with the Heaviside theta function $\Theta$ (see fig.\ \ref{fig:step}).  In the limit $\alpha \to \infty$, the Woods-Saxon potential reaches the Heaviside step potential.

%%%%%%%%%%%%%%%%%%%%%%%%%%%%%%%
\subsection{Eigenstates of the smooth step}
The time-independent Schrödinger equation 
\begin{align}
    \hat{H} \phi_k(x) = \frac{k^2}{2m}\phi_k(x)\,, \label{eq:TS}
\end{align}
of the smooth step $V_{WS}$, with the energy $E=k^2/(2m)$, is solved in closed form using the Gaussian or ordinary hypergeometric function 
\begin{align}
    \tensor[_2]{F}{_1}(a, b; c; z) = \frac{\Gamma(c)}{\Gamma(a) \Gamma(b)}\sum_{n=0}^\infty \frac{\Gamma(a+n) \Gamma(b+n)}{\Gamma(c+n)}\frac{z^n}{n!}\,,
\end{align}
on the real interval $-1<z<1$ and its analytic continuation. We solve the time-independent Schrödinger equation for energies above and below the height of the step.
\begin{widetext}
\begin{itemize}
    \item When the energy $E=\frac{k^2}{2m}$ is below step strength $V_0$, there exists one physical eigenstate
    \begin{align}
        \phi^c_k(x) &= 2^{-\frac{i k + \mu}{2 \alpha \hbar}} e^{\frac{(ik- \mu)x}{2 \hbar}} \text{sech}( \alpha x)^{\frac{i k+ \mu}{2 \alpha \hbar}}\,
        \, \tensor[_2]{F}{_1}\left(1 + \frac{ik+ \mu}{2 \alpha \hbar}, \frac{i k + \mu}{2 \alpha \hbar}; 1 +\frac{ \mu}{ \alpha \hbar}; \frac{1}{1+e^{2 \alpha x}}\right)\,,
    \end{align}
    with the constant $\mu= \sqrt{2m V_0 - k^2}$. The solution decays exponentially in the classically forbidden region. The second independent solution of the time-independent Schrödinger equation \eqref{eq:TS} diverges in the classically forbidden region and is non-physical.
    \item When the energy $E=\frac{k^2}{2m}$ exceeds the step strength $V_0$, there exist two physical eigenstates
    \begin{align}
        \phi^+_k(x) &= 2^{-\frac{i(k-p)}{2 \alpha \hbar}}e^{\frac{i(k+p)x}{2 \hbar}} \text{sech}( \alpha x)^{\frac{i(k-p)}{2 \alpha \hbar}}\,
        \, \tensor[_2]{F}{_1}\left(1 + \frac{i(k-p)}{2 \alpha \hbar}, \frac{i(k-p)}{2 \alpha \hbar}; 1 -\frac{i p}{ \alpha \hbar}; \frac{1}{1+e^{2 \alpha x}}\right)\,,\\
        %%%%%%%%%%%%%%%%%%%%
        \phi^-_k(x) &= 2^{-\frac{i(k+p)}{2 \alpha \hbar}}e^{\frac{i(k-p)x}{2 \hbar}} \text{sech}( \alpha x)^{\frac{i(k+p)}{2  \alpha \hbar}}\, 
        \, \tensor[_2]{F}{_1}\left(1 + \frac{i(k+p)}{2 \alpha \hbar}, \frac{i(k+p)}{2 \alpha \hbar}; 1 +\frac{i p}{ \alpha \hbar}; \frac{1}{1+e^{2 \alpha x}}\right)\,,
\label{eqn:tanheig}
    \end{align}
    with the magnitude of the momentum of a particle on the right of the step $p=\sqrt{k^2 - 2m V_0}$. The two eigenstates are related by the exchange $p \leftrightarrow -p$. 
\end{itemize}
\end{widetext}
\begin{figure*}
    \centering
    \begin{subfigure}[b]{0.49\textwidth}
        \centering
        \includegraphics[width=\textwidth]{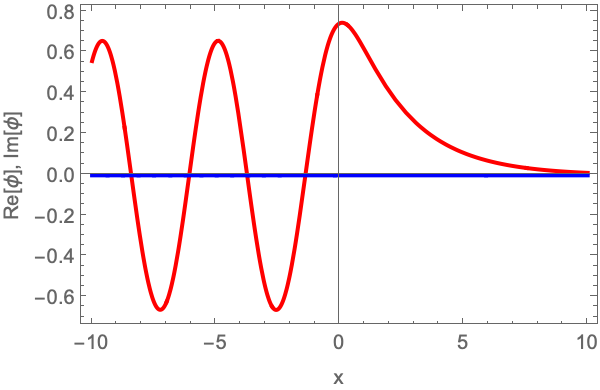}
        \caption{$\phi^c_k(x)$ for $k=0.95 \sqrt{2 m V_0}$}
    \end{subfigure}
    \hfill
    \begin{subfigure}[b]{0.49\textwidth}
        \centering
        \includegraphics[width=\textwidth]{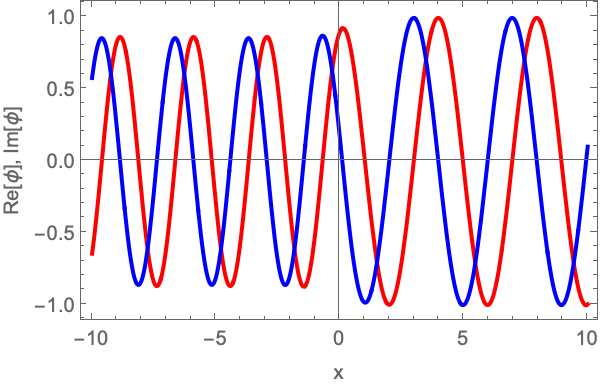}
        \caption{$\phi^+_k(x)$ for $k=1.5 \sqrt{2 m V_0}$}
    \end{subfigure}
    \caption{The real (red) and imaginary part (blue) of the eigenstates $\phi^c$, $\phi^+$ for $m=1,V_0=1$, and $\alpha=1$.}\label{fig:eigenstates}
\end{figure*}
These eigenstates are equivalent to the ones presented in \cite{Birrell:1984} and \cite{Grosche:1989}. For energies below the height of the step, the eigenstates $\phi_k^c$ oscillate on the left and exponentially decay as an evanescent wave in the classically forbidden region as $x \to \infty$ (see the left panel of fig.\ \ref{fig:eigenstates}). For energies above the height of the step, the eigenstates $\phi_k^\pm$ oscillate on the left and the right of the step with different amplitudes (see the right panel of fig.\ \ref{fig:eigenstates}). The eigenstates $\phi_k^-$ and $\phi_k^+$ show qualitatively similar behaviour. For clarity, we will only consider real positive $k, \mu$ and $p$ corresponding to the magnitude of the momenta. Note that the eigenstates are preserved under the scaling $\alpha \mapsto A \alpha$, $x \mapsto A^{-1} x$ and $\hbar \mapsto A^{-1} \hbar$ for a nonzero constant $A$.

\begin{figure}
    \centering
    \includegraphics[width=\linewidth]{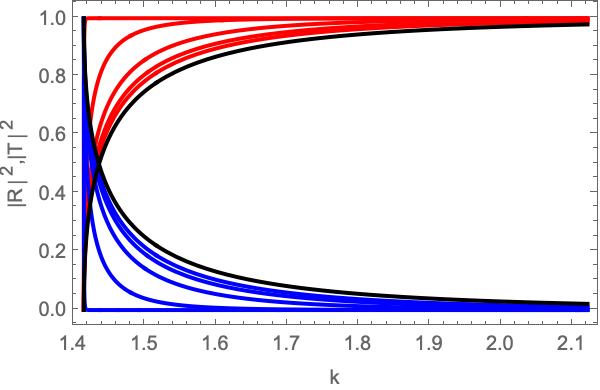}
    \caption{The reflection and transmission rates $|\tilde{R}_k|^2$ (blue) and $|\tilde{T}_k|^2$ (red) for $m=1,V_0=1,\hbar=1$ for $\alpha = 0.1, 1, 2, 3, 4$. The reflection and transmission rates for the Heaviside step potential are plotted in black.}\label{fig:TR}
\end{figure}

%%%%%%%%%%%%%%%%%%%%%%%%%%%%%%%
\begin{widetext}
\subsection{Asymptotics of the smooth step}\label{sec:asymptotics}
In the limits $x\to \pm \infty$, the particle is free and the eigenstates approach plane waves:
\begin{itemize}
    \item For energies below the step, the eigenfunction approaches the asymptotics
    \begin{align}
        \phi^c_k(x) &\underset{x \to -\infty}{\sim}\frac{ \pi  \Gamma\left(1 + \frac{\mu}{\alpha \hbar}\right)}{2 \alpha \hbar \sinh \frac{\pi k}{\alpha \hbar}}
    \left[
        \frac{(k-i \mu) e^{-\frac{i k x}{\hbar}}}{\Gamma\left(1 - \frac{ i k }{\alpha \hbar}\right) \Gamma\left(1+ \frac{i k+ \mu}{2 \alpha \hbar}\right)^2} 
        +
        \frac{(k+i \mu) e^{\frac{i k x}{\hbar}}}{\Gamma\left(1 + \frac{ i k }{\alpha \hbar}\right) \Gamma\left(1- \frac{i k- \mu}{2 \alpha \hbar}\right)^2}\right]\,,    
    \end{align}
    and 
    \begin{align}
        \phi^c_k(x) \underset{x \to \infty}{\sim}e^{-\frac{\mu x}{\hbar}}\,.
    \end{align}
    \item For energies above the step, the eigenfunctions approach the form, 
    \begin{align}
        \phi^-_k(x) &\underset{x \to -\infty}{\sim} \frac{ \pi  \Gamma\left(1 + \frac{ i p}{\alpha \hbar}\right)}{2 \alpha \hbar \sinh \frac{\pi k}{\alpha \hbar}}
        \left[
            \frac{(k+p) e^{-\frac{i k x}{\hbar}}}{\Gamma\left(1 - \frac{ i k }{\alpha \hbar}\right) \Gamma\left(1+ \frac{i (k+p)}{2 \alpha \hbar}\right)^2} 
            +
            \frac{(k-p) e^{\frac{i k x}{\hbar}}}{\Gamma\left(1 + \frac{ i k }{\alpha \hbar}\right) \Gamma\left(1- \frac{i (k-p)}{2 \alpha \hbar}\right)^2}\right]\,,
            \label{eqn:out}\\
            \phi^-_k(x) &\underset{x \to \infty}{\sim}  e^{-\frac{i p x}{\hbar}}\,,
\label{eqn:in}    \end{align}
    and
    \begin{align}
        \phi^+_k(x) &\underset{x \to -\infty}{\sim} \frac{ \pi  \Gamma\left(1 - \frac{ i p}{\alpha \hbar}\right)}{2 \alpha \hbar \sinh \frac{\pi k}{\alpha \hbar}}
        \left[
            \frac{(k-p) e^{-\frac{i k x}{\hbar}}}{\Gamma\left(1 - \frac{ i k }{\alpha \hbar}\right) \Gamma\left(1+ \frac{i (k-p)}{2 \alpha \hbar}\right)^2} 
            +
            \frac{(k+p) e^{\frac{i k x}{\hbar}}}{\Gamma\left(1 + \frac{ i k }{\alpha \hbar}\right) \Gamma\left(1- \frac{i (k+p)}{2 \alpha \hbar}\right)^2}\right]\,,\\
            \phi^+_k(x) &\underset{x \to \infty}{\sim}  e^{\frac{i p x}{\hbar}}\,.
    \end{align}
\end{itemize}
\end{widetext}
By identifying the asymptotics of the eigenstates above the step $\phi_k^\pm$ with the parametrisation 
\begin{align}
    \phi^+_k(x) &\underset{x \to -\infty}{\sim} \mathcal{N}_k\left( \tilde{R}_k e^{-ik x/\hbar} + e^{ik x/\hbar}\right)\,,\\
    \phi^+_k(x) &\underset{x \to +\infty}{\sim} \mathcal{N}_k\, \tilde{T}_k \sqrt{ k/p}e^{ip x/\hbar}\,,
\end{align}
we obtain the reflection and transmission amplitudes
\begin{align}
    \tilde{R}_k &= 
    \frac{k-p}{k+p}
    \frac{\Gamma\left(1 + \frac{ i k }{\alpha \hbar}\right) \Gamma\left(1- \frac{i (k+p)}{2 \alpha \hbar}\right)^2}{\Gamma\left(1 - \frac{ i k }{\alpha \hbar}\right) \Gamma\left(1+ \frac{i (k-p)}{2 \alpha \hbar}\right)^2}\,, \\
    \tilde{T}_k &=
    \frac{2 \alpha \hbar \sinh \frac{\pi k}{\alpha \hbar}\sqrt{p/k}}{k+p}
    \frac{\Gamma\left(1 + \frac{ i k }{\alpha \hbar}\right) \Gamma\left(1- \frac{i (k+p)}{2 \alpha \hbar}\right)^2}{ \pi  \Gamma\left(1 - \frac{ i p}{\alpha \hbar}\right)}\,,
\end{align}
and the corresponding reflection and transmission rates
\begin{align}
    |\tilde{R}_k|^2 = \frac{ \sinh^2 \frac{\pi(k-p)}{2 \alpha \hbar}}{\sinh^2 \frac{ \pi(k+p)}{2 \alpha \hbar}}\,,\quad
    |\tilde{T}_k|^2 =\frac{ \sinh \frac{ \pi k}{\alpha \hbar} \sinh\frac{\pi p}{\alpha \hbar}}{\sinh^2 \frac{\pi(k+p)}{2 \alpha \hbar}}\,.
\label{eqn:RT}
\end{align}
See fig.\ \ref{fig:TR} for an illustration of the reflection and transmission rates. In the limit $\hbar \to 0$ for fixed $\alpha>0$, the rate reflection rate approaches the asymptotic
\begin{align}
    |\tilde{R}_k|^2 \underset{\hbar \to 0}{\sim} e^{-\frac{2 \pi p}{\alpha \hbar}} = e^{\frac{2 i S_I}{\hbar}}\,,
\end{align} 
with an ``instanton''-like action $S_I=\pi i \sqrt{2m(E-V_0)}/\alpha$.  This paper will illuminate the classical dynamics of $S_I$.

For large incident momenta, the transmission rate $|\tilde{T}_k|^2$ approaches unity. For a smooth step, a particle coming from the left with kinetic energy exceeding the step continues its journey to the right, irrespective of the steepness of the step. On the other hand, in the limit $\alpha \to \infty$ reaching the Heaviside step potential for fixed $\hbar>0$, we obtain the reflection $|\tilde{R}_k|^2 \to \frac{(k-p)^2}{(k+p)^2}$ and transmission rate $|\tilde{T}_k|^2 \to \frac{4 k p}{(k+p)^2}$. The Heaviside step potential $V_H$ is reflective irrespective of the Planck constant. This is, of course, not a physical prediction as physical steps can only approach the Heaviside step potential and will include a boundary layer. For the step function limit, the reflection is not exponentially suppressed for small $\hbar$, seemingly at odds with an instanton picture.

%%%%%%%%%%%%%%%%%%%%%%%%%%%%%%%
\subsection{The Feynman path integral}
We express the Feynman path integral in terms of the eigenstates using the spectral representation
\begin{align}
    &G(x_1,x_0;T) = \Theta(T) \bigg[
    \int_0^{\sqrt{2 m V_0}}\varphi^c_k(x_1) \varphi^c_k(x_0)^* e^{- \frac{i k^2T}{2m \hbar} }\mathrm{d}k \nonumber\\
    &+\int_{\sqrt{2 m V_0}}^\infty \left(\varphi^-_k(x_1) \varphi^-_k(x_0)^* + \varphi^+_k(x_1) \varphi^+_k(x_0)^*\right) e^{- \frac{i k^2T}{2m \hbar} }\mathrm{d}k
    \bigg]\,,
    \label{eq:spectral}
\end{align}
with the orthonormal eigenstates $\varphi_k^c,\varphi_k^\pm$ satisfying the time-independent Schrödinger equation and the normalization
\begin{align}
    \int \varphi^a_k(x) \varphi^b_{k'}(x)^*\mathrm{d}x=\delta_{ab}\delta\left(\frac{k-k'}{\hbar}\right)\,,
\end{align}
with the Kronecker delta $\delta_{ab}$ running over $a,b = c,+$ and $-$. The energy propagator, describing the transition amplitude between two states $x_0$ and $x_1$ at energy $E$, is defined as the Fourier transform of the real-time propagator
\begin{align}
    K(x_1,x_0;E) = \int_0^\infty G(x_1,x_0;T) e^{i E T/ \hbar}\mathrm{d}T\,,
\end{align}
leading to the spectral representation
\begin{align}
    &K(x_1,x_0;E) =
    i \hbar  \bigg[
    \int_0^{\sqrt{2 m V_0}}\frac{\varphi^c_k(x_1) \varphi^c_k(x_0)^* }{E - \frac{k^2}{2m}} \mathrm{d}k\nonumber\\
    &+\int_{\sqrt{2 m V_0}}^\infty \frac{\varphi^-_k(x_1) \varphi^-_k(x_0)^* + \varphi^+_k(x_1) \varphi^+_k(x_0)^* }{E - \frac{k^2}{2m}} \mathrm{d}k
    \bigg]\,.\label{eq:spectral_energy}
\end{align}

To construct the orthonormal eigenstates $\varphi_k^c,\varphi^\pm_k$, we evaluate the inner products of the eigenstates $\phi_k^c,$ and $\phi_k^\pm$ described above. The spectral method has been known for a long time, see for example  \cite{Feynman:1948, Griffiths:2018}. However, to the best of our knowledge, the here proposed derivation of the overlap of eigenstates has not previously been presented in the present form. This derivation extends to any quantum system with known eigenstates.

\bigskip
As the Hamiltonian is Hermitian, the eigenstates of different energies are orthogonal, \textit{i.e.}, for the eigenstates $\phi_k^a$ and $\phi_{k'}^b$ with $k\neq k'$,
\begin{align}
    \int_{-\infty}^\infty \phi_k^a(x) \phi_{k'}^b(x)^* \mathrm{d}x=0\,.
\end{align}
Consequently, we focus on the case for which $k=k'$. Using the time-independent Schrödinger equation 
\begin{align}
    &\left[-\frac{\hbar^2}{2m}\frac{\partial^2}{\partial x^2} + V(x)\right] \phi_k^a(x) = \frac{k^2}{2m}\phi_k^a(x)\,,\\
    &\left[-\frac{\hbar^2}{2m}\frac{\partial^2}{\partial x^2} + V(x)\right] \phi_{k'}^b(x)^* = \frac{k'^2}{2m}\phi_{k'}^b(x)^*\,,
\end{align}
for general $k$ and $k'$, we infer the overlap for the case $k=k'$. Upon multiplying the first and second equations by $\phi_{k'}^b(x)^*$ and $\phi_k^a(x)$ respectively, and subtracting the resulting equations, the potential term drops out, yielding the identity
\begin{align}
    \frac{\partial^2 \phi_k^a}{\partial x^2}\phi_{k'}^{b*} - \phi_k^a\frac{\partial^2 \phi_{k'}^{b*}}{\partial x^2} = \frac{k'^2 -k^2}{\hbar^2} \phi_k^a \phi_{k'}^{b*} \,.
\end{align}
Integrating this identity over the domain $[A, B]$, we find upon integrating by parts
\begin{align}
   \int_A^B \phi_k^a(x) \phi_{k'}^b(x)^* \mathrm{d}x 
   &=
   \frac{\hbar^2}{k'^2 -k^2}
   \left[\frac{\partial \phi_k^a}{\partial x} \phi^{b*} - \phi^a \frac{\partial \phi_{k'}^{b*}}{\partial x}\right]_A^B\,.
\end{align}
Remarkably, the time-independent Schrödinger equation allows us to write the integral of the product of eigenstates as their values at the boundary points. In the limit, $A\to -\infty$ and $B\to \infty$, the inner product assumes the form
\begin{align}
    \int_{-\infty}^\infty \phi_k^a(x) \phi_{k'}^b(x)^* \mathrm{d}x 
    &=
    \frac{\hbar^2}{k'^2 -k^2}
    \left[\frac{\partial \phi_k^a}{\partial x} \phi_{k'}^{b*} - \phi_k^a \frac{\partial \phi_{k'}^{b*}}{\partial x}\right]_{-\infty}^\infty\,.\label{eq:identity1}
\end{align}
The overlap of the two states is solely determined by the asymptotic behaviour of the eigenstates and does not explicitly depend on the potential. At first sight, the case of interest $k = k'$ seems ill-defined as we divide by the square of the difference. However, as we will see, the pole is cancelled by the oscillatory nature of the limits $A\to -\infty$ and $B \to \infty$, making the limit $k' \to k$ well-defined. This identity applies to any potential $V$.

Assuming the particle to be free for large $|x|$ and satisfying the asymptotics
\begin{align}
    \phi_k^a(x) &\sim A_k e^{-ik x/\hbar} + B_k e^{ik x/\hbar}\,, \\
    \phi_{k'}^b(x) &\sim C_{k'} e^{-ik'x/\hbar} + D_{k'} e^{ik'x/\hbar}\,,
\end{align}
for some factors $A_k, B_k, C_{k'}$ and $D_{k'}$ in the limit $x\to -\infty$ or $x \to \infty$ (with not necessarily the same factors in the two limits), we obtain the identity
\begin{align}
    &\frac{\hbar^2}{k'^2 -k^2}
    \lim_{x \to \pm \infty}\left[\frac{\partial \phi_k^a}{\partial x} \phi_{k'}^{b*} - \phi_k^a \frac{\partial \phi_{k'}^{b*}}{\partial x}\right] \nonumber\\
    &= \pm \pi \left(A_k C_{k}^* + B_k D_{k}^*\right) \delta\left(\frac{k-k'}{\hbar}\right)\,,
\end{align}
using the limit 
\begin{align}
    \lim_{x \to \infty} \frac{e^{i k x}}{k} = i \pi \delta(k),  
\end{align}
which explains the occurrence of the pole at $k=k'$ in equation \eqref{eq:identity1}. Upon applying the asymptotics of the eigenstates (see section \ref{sec:asymptotics}), we obtain the inner products
\begin{align}
    \int_{-\infty}^\infty \phi_k^c(x) \phi_{k'}^c(x)^*\mathrm{d}x
    &= \mathcal{N}^{cc}_k \delta\left(\frac{k-k'}{\hbar}\right)\,,\\
    \int_{-\infty}^\infty \phi_k^+(x) \phi_{k'}^+(x)^*\mathrm{d}x &= 
    \int_{-\infty}^\infty \phi_k^-(x) \phi_{k'}^-(x)^*\mathrm{d}x \nonumber\\
    &= \mathcal{N}^{++}_k \delta\left(\frac{k - k'}{\hbar}\right)\,,\\
    \int_{-\infty}^\infty \phi_k^+(x) \phi_{k'}^-(x)^*\mathrm{d}x
    &=\mathcal{N}^{+-}_k 
    \delta\left(\frac{k-k'}{\hbar}\right)\,,
\end{align}
with the coefficients
\begin{widetext}
\begin{align}
    &\mathcal{N}^{cc}_k  
    = \frac{ m V_0 }{ \alpha \hbar k \sinh \frac{\pi k}{\alpha \hbar}} \frac{\pi^2\Gamma\left(1+ \frac{\mu}{\alpha \hbar}\right)^2}{\left|\Gamma\left(1+ \frac{\mu + i k}{2 \alpha \hbar}\right)\right|^4}\,,\\
    &\mathcal{N}^{++}_k = \frac{\pi p}{k} \left[1
    +\frac
{\left(\cosh \frac{\pi k}{\alpha \hbar} - \cosh \frac{\pi p}{\alpha \hbar}\right)^2  \left(\csch^2 \frac{\pi (k-p) }{2 \alpha \hbar} + \csch^2\frac{\pi (k+p)}{2 \alpha \hbar }\right)}{4\sinh\frac{\pi k }{\alpha \hbar} \sinh \frac{ \pi p}{\alpha \hbar}}
    \right]\,,\\
    &\mathcal{N}^{+-}_k 
    =\frac{ m V_0}{ \alpha \hbar k \sinh \frac{ \pi k}{\alpha \hbar} }\frac{ \pi^2 \Gamma\left(1- \frac{i p}{\alpha \hbar}\right)^2}{ \Gamma\left(1+ \frac{i(k-p)}{2 \alpha \hbar}\right)^2 \Gamma\left(1 - \frac{i (k+p)}{2 \alpha \hbar}\right)^2}\,.
\end{align}
\end{widetext}
Using these inner products, we construct an orthonormal basis of eigenstates. When the energy is below the height of the step, we obtain the state
\begin{align}
    \varphi^c_k(x) = \frac{\phi^c_k(x)}{\sqrt{\hbar \mathcal{N}^{cc}_k}}\,,
\end{align}
ensuring that $\int \varphi^c_k(x) \varphi^c_{k'}(x)^*\mathrm{d}x=\delta\left(\frac{k-k'}{\hbar}\right)$.
When the energy is above the height of the step, we define $\varphi^\pm$ as
% \begin{align}
%     \varphi^+_k(x) &= \frac{\phi^+_k(x)}{\sqrt{\hbar \mathcal{N}^{++}_k }}\\
%     \varphi^-_k(x) &= \frac{(\mathcal{N}^{+-}_k )^* \phi^+_k(x) - \mathcal{N}^{++}_k \phi^-_k(x)}{\sqrt{\hbar \mathcal{N}^{++}_k ((\mathcal{N}^{++}_k )^2-|\mathcal{N}^{+-}_k |^2)}}
% \end{align}
\begin{align}
    \varphi^+_k(x) &= \frac{\phi_k^+(x) + \frac{\mathcal{N}_k^{+-}}{|\mathcal{N}_k^{+-}|}\phi_k^-(x)}{\sqrt{2 \hbar (\mathcal{N}_k^{++}+|\mathcal{N}_k^{+-}|)}}\\
    \varphi^-_k(x)  &= \frac{\phi_k^+(x) - \frac{\mathcal{N}_k^{+-}}{|\mathcal{N}_k^{+-}|}\phi_k^-(x)}{\sqrt{2 \hbar (\mathcal{N}_k^{++} - |\mathcal{N}_k^{+-}|)}}
\end{align}
ensuring that $\int \varphi^a_k(x) \varphi^{b}_{k'}(x)^*\mathrm{d}x=\delta_{ab}\delta\left(\frac{k-k'}{\hbar}\right)$ with $a,b=\pm$. Evaluation of the terms in the numerator and denominator shows
% \begin{widetext}
\begin{align}
    \frac{\mathcal{N}_k^{+-}}{|\mathcal{N}_k^{+-}|} 
    = \frac{\Gamma\left(1 - \frac{i (k-p)}{2 \alpha \hbar}\right) \Gamma\left(1 - \frac{i p}{\alpha \hbar}\right) \Gamma\left(1+ \frac{i (k+p)}{2 \alpha \hbar}\right)}{\Gamma\left(1+ \frac{i(k-p)}{2 \alpha \hbar}\right)\Gamma\left(1+\frac{i p}{\alpha \hbar}\right) \Gamma\left(1 - \frac{i (k+p)}{2 \alpha \hbar}\right)}\,,
\end{align}
and 
\begin{align}
    \mathcal{N}_k^{++} + |\mathcal{N}_k^{+-}| = \frac{2 \pi p}{k} \frac{1 - e^{\frac{\pi(k+p)}{\alpha \hbar}}}{\left(1+e^{\frac{\pi k }{\alpha \hbar}}\right)\left(1-e^{\frac{\pi p }{\alpha \hbar}}\right)}\,,\\
    \mathcal{N}_k^{++} - |\mathcal{N}_k^{+-}| = \frac{2 \pi p}{k} \frac{1 - e^{\frac{\pi(k+p)}{\alpha \hbar}}}{\left(1-e^{\frac{\pi k }{\alpha \hbar}}\right)\left(1+e^{\frac{\pi p }{\alpha \hbar}}\right)}\,.
\end{align}
% \end{widetext}

\begin{figure*}
    \centering
    \begin{subfigure}[b]{0.32\linewidth}
        \includegraphics[width=\textwidth]{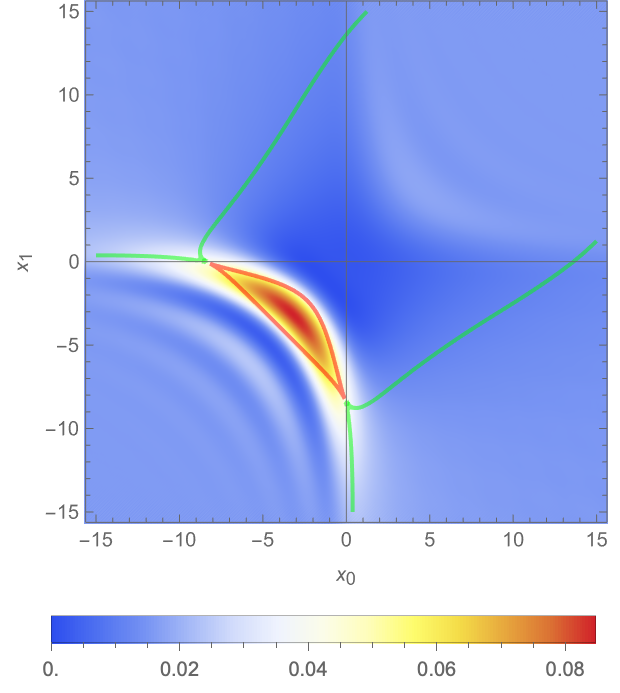}
        \caption{$\hbar=1$}
    \end{subfigure}
    \hfill
    \begin{subfigure}[b]{0.32\linewidth}
        \includegraphics[width=\textwidth]{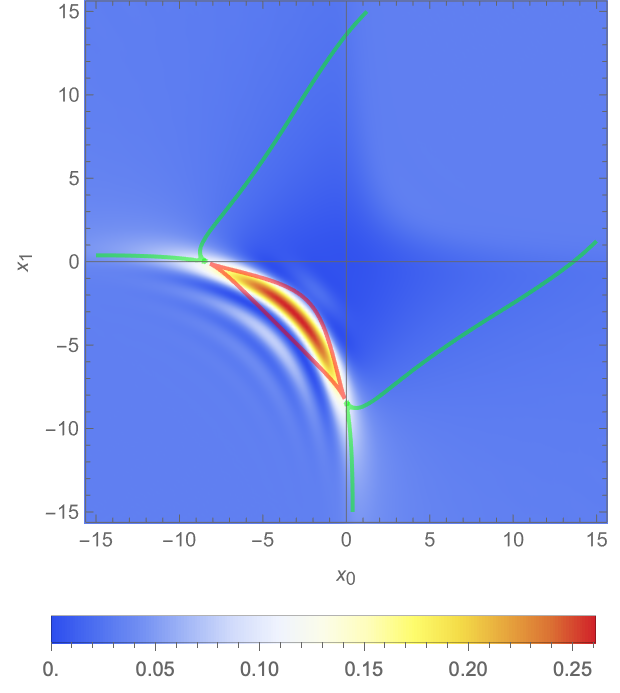}
        \caption{$\hbar=0.5$}
    \end{subfigure}
    \hfill
    \begin{subfigure}[b]{0.32\linewidth}
        \includegraphics[width=\textwidth]{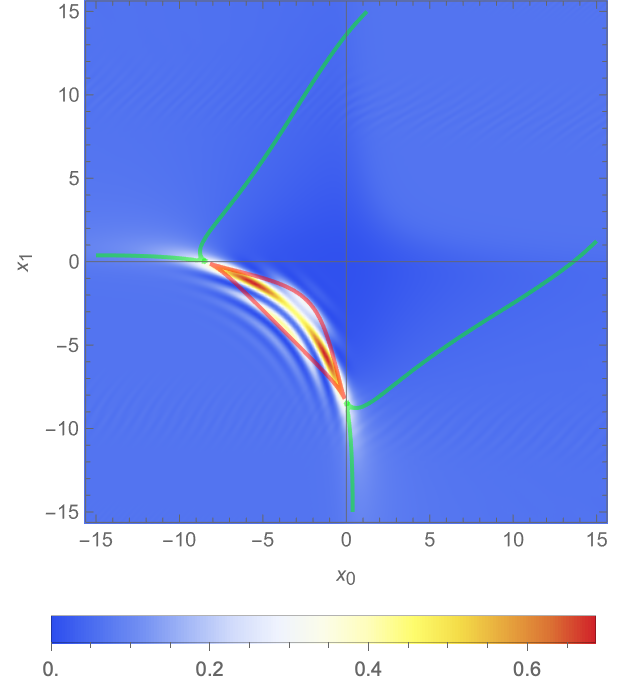}
        \caption{$\hbar=0.25$}
    \end{subfigure}
    \caption{The real-time propagator $|G(x_1,x_0;T)|^2$ for the smooth step for $\hbar=1, 0.5$ and $0.25$ (left to right) for $T=10, m=1, \alpha=1$ and $V_0=1$. The red curves mark the caustics. The green arcs mark the Stokes lines.}\label{fig:Tanh_Propagator}
\end{figure*}

Upon substituting the orthonormal eigenstates in the spectral representation \eqref{eq:spectral}, we reduce the evaluation of the infinite-dimensional path integral to a one-dimensional integral over the magnitude of the momentum $k$. We evaluate this highly oscillatory integral numerically by deforming the integration domain of the second integral slightly into the complex plane $[\sqrt{2 m V_0},\infty) \mapsto[\sqrt{2 m V_0},\infty  e^{-i \theta})$ for a small positive deformation angle $\theta$. This deformation, inspired by the more general Picard-Lefschetz theory \cite{Feldbrugge:2023d}, removes the oscillatory nature of the integrand and ensures that the integrand decays exponentially away from the point $\sqrt{2 m V_0}$. For large negative $\alpha x_0$ or $\alpha x_1$, the evaluation of the propagator in the current form is numerically unstable, as the argument of the hypergeometric function $1/(1+e^{2 \alpha x})$ exponentially approaches the pole of the hypergeometric function at $z = 1$ and becomes sensitive to numerical noise. To bypass this problem, we replace the orthogonal eigenstates by their asymptotics in the spectral representation for large $|\alpha x_0|$ and $|\alpha x_1|$.

The resulting spectral representation agrees with the one presented in \cite{Grosche:1989}. The path integral evaluates to an interference pattern in $x_0$-$x_1$-$T$ space (see fig.\ \ref{fig:Tanh_Propagator}) not unlike the lensing patterns emerging in wave optics \cite{Nye:1974, Nye:1999, Feldbrugge:2023d}. As we will see in the next section, the nature of the interference pattern can be understood in terms of solutions of the classical boundary value problem. The interference pattern becomes more oscillatory as we decrease the Planck constant and approach the semi-classical limit.

%%%%%%%%%%%%%%%%%%%%%%%%%%%%%%%
\section{The Heaviside step}\label{sec:Heaviside_step}
In the limit $\alpha \to \infty$, we recover the real-time path integral for a particle interacting with the Heaviside step potential $V_H$. 
\begin{itemize}
    \item For energies below the height of the step, we obtain the eigenstate
    \begin{align}
        \phi_k^c(x) &= \begin{cases}
            \frac{k-i \mu}{2k} e^{-ikx / \hbar} + \frac{k+i \mu}{2k} e^{i k x /\hbar}, & x \leq 0\,,\\
           e^{- \mu x / \hbar}, & x  > 0\,.
        \end{cases}
    \end{align}
    \item For energies above the step, we obtain
    \begin{align}
        \phi_k^+(x) &= \begin{cases}
            \frac{k-p}{2k} e^{-ikx / \hbar} + \frac{k+p}{2k} e^{i k x /\hbar}, & x \leq 0\,,\\
           e^{i p x / \hbar}, & x  > 0\,,
        \end{cases}
    \end{align}
    and
    \begin{align}
         \phi_k^-(x) &= \begin{cases}
            \frac{k+p}{2k} e^{-ikx / \hbar} + \frac{k-p}{2k} e^{i k x /\hbar} , & x \leq 0\,,\\
            e^{-i p x / \hbar}, & x  > 0\,.
         \end{cases}
    \end{align}
\end{itemize}

The asymptotics yield the well-known reflection and transmission amplitudes 
\begin{align}
    \tilde{R}_k = \frac{k-p}{k+p}\,, \quad
    \tilde{T}_k = \frac{2 \sqrt{k p}}{k+p}\,.
\end{align}
The modulus squared $|\tilde{R}_k|^2$ and $|\tilde{T}_k|^2$ are the corresponding reflection and transmission rates.

The inner products of the eigenstates 
\begin{align}
    \mathcal{N}^{cc}_k &= \mathcal{N}^{+-}_k = \pi \frac{m V_0}{k^2} \,,\\
    \mathcal{N}^{++}_k &=\pi \left[\frac{p}{k} + \frac{k^2 + p^2}{2 k^2}\right]\,,
\end{align}
lead to the orthonormal eigenstates 
\begin{align}
    \varphi_k^c(x) &= \frac{1}{\sqrt{\pi \hbar m V_0 }} \begin{cases}
        k\cos \frac{k x}{\hbar} -  \mu \sin \frac{k x}{\hbar}, & x \leq 0\,,\\
        k e^{- \mu x / \hbar}, & x  > 0\,,
    \end{cases}\\
    \varphi_k^+(x) &= \frac{2 }{\sqrt{\pi \hbar ((k+p)^2+2 m V_0)}}
    \begin{cases}
       k \cos \frac{k x}{\hbar}, & x \leq 0\,,\\
       k \cos \frac{p x}{\hbar}, & x  > 0\,,
    \end{cases}\\
    \varphi_k^-(x) &= \frac{2i}{\sqrt{\pi \hbar ((k+p)^2-2 m V_0)}}
    \begin{cases}
        p \sin \frac{k x}{\hbar}, & x \leq 0\,,\\
        k \sin \frac{p x}{\hbar}, & x  > 0\,.
    \end{cases}
\end{align}

\begin{figure*}
    \centering
    \begin{subfigure}[b]{0.32\linewidth}
        \includegraphics[width=\textwidth]{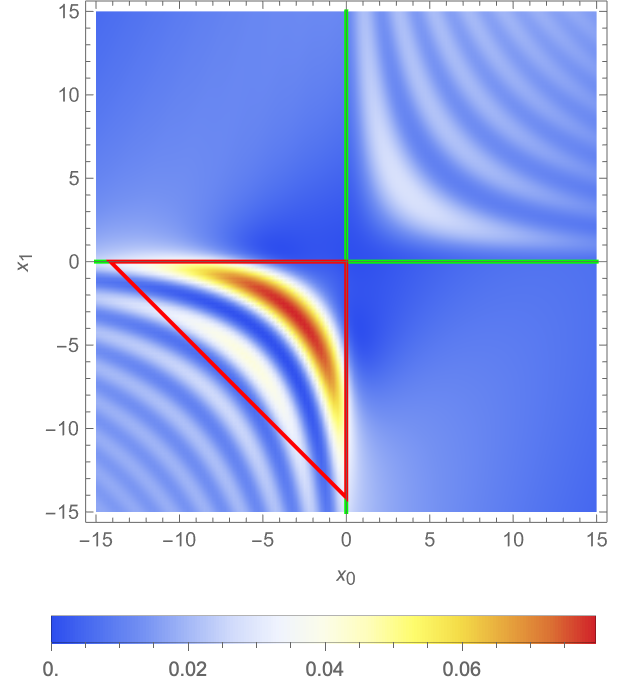}
        \caption{$\hbar=1$}
    \end{subfigure}
    \hfill
    \begin{subfigure}[b]{0.32\linewidth}
        \includegraphics[width=\textwidth]{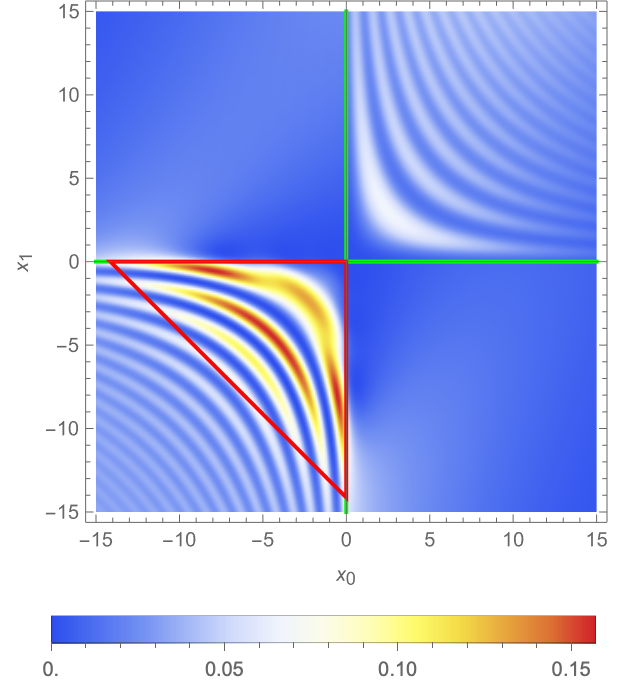}
        \caption{$\hbar=0.5$}
    \end{subfigure}
    \hfill
    \begin{subfigure}[b]{0.32\linewidth}
        \includegraphics[width=\textwidth]{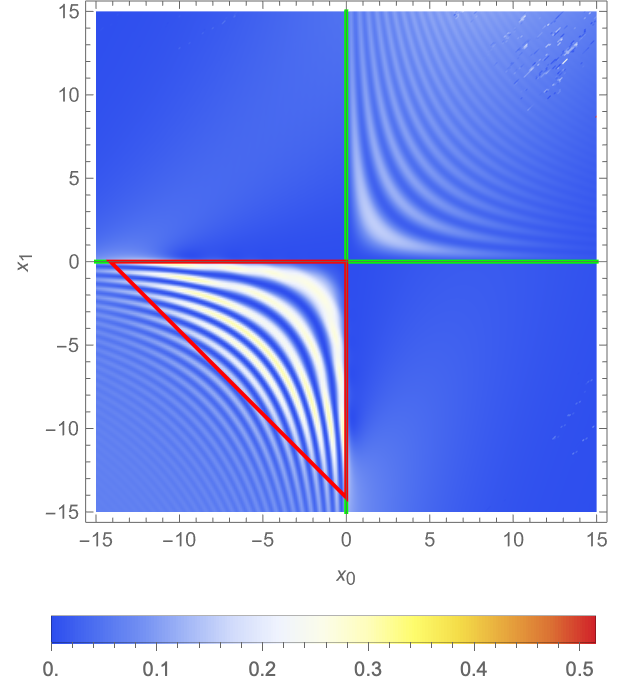}
        \caption{$\hbar=0.25$}
    \end{subfigure}
    \caption{The real-time propagator $|G(x_1,x_0;T)|^2$ for the Heaviside step for $\hbar=1,0.5,$ and $0.25$ (left to right) for $T=10, m=1$ and $V_0=1$. The red lines mark the caustics. The green lines mark the Stokes lines.}\label{fig:Step_Propagator}
\end{figure*}

The real-time and energy propagators are obtained with the spectral representations \eqref{eq:spectral} and \eqref{eq:spectral_energy}, consistent with \cite{Aguiar:1993}. To improve the performance of the numerical evaluation of the integral, we deform the integration domain $[\sqrt{2m V_0},\infty)$ into the complex plane. For fixed propagation time $T$, the path integral consists of an interference pattern in the $x_0$-$x_1$ plane (see fig.\ \ref{fig:Step_Propagator}). Note that the steeper step has a larger domain for which three real classical paths solve the classical boundary value problem. As we decrease the Planck constant $\hbar$, the interference pattern becomes increasingly oscillatory.

%%%%%%%%%%%%%%%%%%%%%%%%%%%%%%%
\section{Classical paths}\label{sec:classical-paths}
The propagator consists of an interference pattern that can be understood in terms of real and complex (semi-) classical paths. As the propagator is a transition amplitude between two states, we will consider solutions of the classical boundary value problem
\begin{align}
    m \ddot{x} = - V'(x)\,, \quad x(0)= x_0\,, \quad x(T)=x_1\,.\label{eq:BVP}
\end{align}
We analyse the real and complex solutions of this boundary value problem, study how they contribute to the real-time Feynman path integral, and conversely, study how the classical theory emerges from the quantum description.

%%%%%%%%%%%%%%%%%%%%%%%%%%%%%%%
\subsection{Real classical paths}\label{sec:real_classical}
\begin{figure*}
    \centering
    \begin{subfigure}[b]{0.49\linewidth}
        \includegraphics[width=\linewidth]{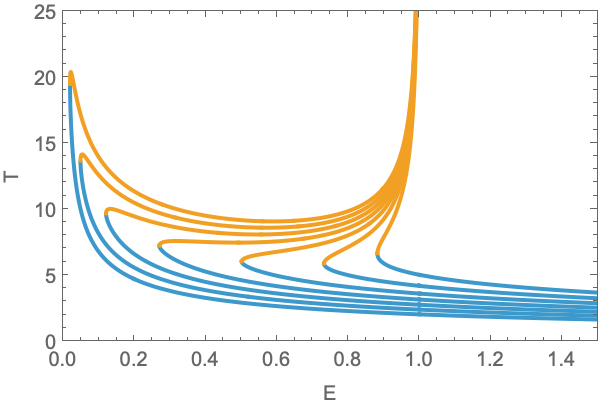}
        \caption{The propagation time \eqref{eq:t1} (blue) and \eqref{eq:t2} (yellow) as a function of the energy $E$, for the initial position $x_0=-5$ and $x_1=-2, -1.5, -1, \dots, 1$ (from left to right) for $V_0=1, \alpha=1$.
        \phantom{Comparison of the paths for a particle starting and ending on the left of the smooth step with $\alpha=1$ (solid curves), $5$ (dashed colored curves) and the Heaviside step (dashed black curves) for the propagation time $T=10$} \phantom{and the boundary values $x_0=-4$ and $x_1=-3$. The direct propagation, low energy bounce, and high} \phantom{energy bounce are plotted in blue, yellow and green.}}\label{fig:t_e}
    \end{subfigure}
    \begin{subfigure}[b]{0.49\linewidth}
        \includegraphics[width=\linewidth]{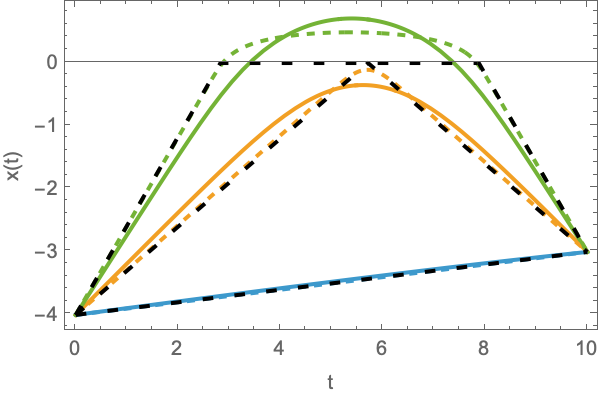}
        \caption{Comparison of the paths for a particle starting and ending on the left of the smooth step with $\alpha=1$ (solid curves), $5$ (dashed colored curves) and the Heaviside step (dashed black curves) for the propagation time $T=10$ and the boundary values $x_0=-4$ and $x_1=-3$. The direct propagation, low energy bounce and high energy bounce are plotted in blue, yellow and green.}
    \label{fig:classicalPaths}
    \end{subfigure}
    \caption{Implicit solutions of the real classical paths. \textit{Left:} the propagation time of the direct and bouncing solution as a function of the energy. \textit{Right:} the direct and bouncing solutions to the boundary value problem \eqref{eq:BVP} for two smooth step potentials and the Heaviside step.}
\end{figure*}

As the potential is independent of time, the classical path conserves energy, 
\begin{align}
    E = \frac{1}{2}m \dot{x}(t)^2 + V(x(t))\,.
\end{align}
The equation
\begin{align}
    \dot{x} = \pm \sqrt{2(E - V(x))/m}\,.
\end{align}
leads to the implicit solution
\begin{widetext}
\begin{align}
    t(x) 
    &= \int \frac{\mathrm{d}x}{\dot{x}}\\
    &=  \pm \int \frac{\mathrm{d}x}{\sqrt{2(E - V(x))/m}} \\
    % &=\frac{\sqrt{m}}{\sqrt{2}\alpha}\left[
    %     \frac{\coth^{-1}\left(\sqrt{\frac{E - V_0}{E - V(x)}}\right)}{\sqrt{E-V_0}}
    %     -
    %     \frac{\coth^{-1}\left(\sqrt{\frac{E }{E - V(x)}}\right)}{\sqrt{E}}\right]\\
    &= \pm \frac{\sqrt{m}}{\sqrt{2}\alpha}\left[
        \frac{1}{\sqrt{E-V_0}}\tanh^{-1}\sqrt{\frac{E - V(x)}{E - V_0}}
        -
        \frac{1}{\sqrt{E}}\tanh^{-1}\sqrt{\frac{E - V(x)}{E
    }}\right] + c_t\,,
\label{eqn:tx}
\end{align}
\end{widetext}
with the integration constant $c_t$ (not yet satisfying the boundary conditions). In this instance, the inverse $x(t)$ cannot be found in closed form. The time has a square root branch point at the turning point
\begin{align}
    x_t = \frac{1}{\alpha} \tanh^{-1}\left(\frac{2 E - V_0}{V_0}\right) \,,\label{eq:turningPoint}
\end{align}
defined by either $\dot{x}=0$ or $E=V(x_t)$. At the turning point $t(x_t)= c_t$, which we will conveniently set to zero ($c_t=0$). We use the implicit solution $t(x)$ to find a solution to the boundary value problem \eqref{eq:BVP}. When the trajectory does not bounce, the propagation time yields
\begin{align}
    T &= \int_{x_0}^{x_1} \frac{\mathrm{d}x}{\sqrt{2 (E-V(x))/m}} \\
    &=  t(x_1) - t(x_0)\,.\label{eq:t1}
\end{align}
Solving for the energy $E$ and shifting the solution such that $t(x_0)=0$, leads to an implicit solution of the boundary value problem \eqref{eq:BVP} (see the blue curves in fig.\ \ref{fig:t_e}).

When the particle reflects off the step, we move around the branch point at the turning point $x_t$, ending on a different Riemann sheet. Consequently, the propagation time assumes the form
\begin{align}
    T &= (t(x_t) - t(x_0)) + (t(x_t) - t(x_1))\\
    &= - (t(x_0) + t(x_1))\label{eq:t2}
\end{align}
See the yellow curves in fig.\, \ref{fig:t_e} for the propagation time $T$ as a function of energy $E$ for different boundary values $x_0,x_1$ at the left of the step. For small propagation times, there exists a single real classical path moving directly between the boundary conditions. For larger propagation times, we observe the creation of bouncing solutions. Note that there are always either one or three real solutions to the boundary value problem. When there exist three real solutions (see fig.\ \ref{fig:classicalPaths}), the path with the lowest energy directly connects the boundary conditions. The two remaining solutions bounce off the step. The bouncing solution with the lowest energy $E$ approaches the step and reflects instantaneously. The second bouncing solution speeds towards the step, with a kinetic energy just below the height of the step $V_0$. The particle lingers in the boundary region before returning to the final position. Remarkably, for some configurations, the path can only reach $x_1$ in the given time $T$ by reflecting off the step. This path corresponds to the classical path that lingers in the boundary layer of the step before returning to $x_1$.

The classical action of the classical path follows from the classical path, \textit{i.e.},
\begin{align}
    S[x] 
    &= \int_0^T \left[\frac{1}{2} m \dot{x}^2 - V(x)\right]\mathrm{d}t\\
    &= \int_0^T \left[m \dot{x}^2 - E \right]\mathrm{d}t\\
    &= -ET +m\int_0^T  \dot{x}^2 \mathrm{d}t\\
    &= -ET +m\int_{x_0}^{x_1}  \dot{x} \mathrm{d}x\\
    &= -ET \pm \int_{x_0}^{x_1}  \sqrt{2m (E-V(x))} \mathrm{d}x\label{eq:Action}
\end{align}    
Defining
\begin{align}
    s(x) &= \int  \sqrt{2m (E-V(x))} \mathrm{d}x\\
    &= \frac{\sqrt{2m}}{\alpha}\bigg[\sqrt{E-V_0} \tanh^{-1}\left(\sqrt{\frac{E-V(x)}{E-V_0}}\right) \nonumber\\
    &\phantom{=}- \sqrt{E} \tanh^{-1}\left(\sqrt{\frac{E-V(x)}{E}}\right)\bigg]\,,\label{eq:sx}
\end{align}
where we have again set the integration constant such that $s$ vanishes at the turning point, \textit{i.e.}, $s(x_t) = 0$ with $E=V(x_t)$, we obtain implicit expressions for the classical action associated with the classical paths.
% 
% \begin{framed}
% \begin{align}
%     S[x] 
%     &= \int_0^T \left[\frac{1}{2} m \dot{x}^2 - V(x)\right]\mathrm{d}t\\
%     &= \int_0^T \left[E -2 V(x)\right]\mathrm{d}t\\
%     &= E T - 2\int_0^T  V(x)\mathrm{d}t\\
%     &= ET - 2 \int_{x_0}^{x_1}  V(x)\frac{\mathrm{d}t}{\mathrm{d}x}\mathrm{d}x\\
%     &= ET \mp 2  \int_{x_0}^{x_1} \frac{ V(x)}{\sqrt{2 (E-V(x))/ m}}\mathrm{d}x\,,
% \end{align}
% using the conservation of energy, with
% \begin{align}
%     s(x) &= \int \frac{ V(x)}{\sqrt{2 (E-V(x))/ m}}\mathrm{d}x\\
%     &=\frac{\sqrt{m} V_0}{\sqrt{2(E-V_0}) \alpha} \tanh^{-1}\sqrt{\frac{E-V(x)}{E-V_0}}\,,
% \end{align}
% \end{framed}
For the direct classical path, we obtain
\begin{align}
    S[x_C] = -ET + s(x_1) -s(x_0)\,.
\end{align}
For the bouncing solution, we find the classical action
\begin{align}
    S[x_C] = -ET - (s(x_0) + s(x_1))\,.\label{eq:S2}
\end{align}

\bigskip
As the smooth step becomes steeper, the bouncing solutions approach the classical paths of the Heaviside step with explicit solutions. We identify three regimes:

\begin{widetext}
    \begin{itemize}
        \item 
    When the initial and final boundary conditions lie on the left side of the step, $x_0,x_1 <0$:
    \begin{itemize}
        \item The linear path assumes the form
        \begin{align}
            x(t) = x_0 + t \frac{x_1-x_0}{T}\,,\label{eq:LL_1}
        \end{align}
        with the corresponding classical action 
        \begin{align}
            S &= \frac{m(x_1-x_0)^2}{2 T}\,.
        \end{align}
        The particle does not experience the step and evolves as a free particle.
        \item The bouncing path that instantaneously reflects of step assumes the form
        \begin{align}
            x(t) = \begin{cases}
                x_0 + t \frac{|x_0| + |x_1|}{T} & \text{ when } t \leq \frac{T |x_0|}{|x_0| + |x_1|}\,,\\
                |x_0| - t \frac{|x_0| + |x_1|}{T} & \text{ when } t \geq \frac{T |x_0|}{|x_0| + |x_1|}\,,
            \end{cases}\label{eq:LL_2}
        \end{align}
        with the corresponding classical action 
        \begin{align}
            S &= \frac{m(x_1+x_0)^2}{2 T}\,.\label{eq:S_2}
        \end{align}
        \item The bouncing path with kinetic $V_0$, lingering at the step for the maximal amount of time, is given by
        \begin{align}
            x(t) = \begin{cases}
                x_0 + \sqrt{2 V_0/ m} t & \text{ when } t \leq \sqrt{\frac{m}{2V_0}}|x_0|\,,\\
                0 & \text{ when } \sqrt{\frac{m}{2V_0}}|x_0| < t < T - \sqrt{\frac{m}{2V_0}}|x_1|\,,\\
                x_1 + \sqrt{2 V_0/m} (T-t) & \text{ when } t \geq T- \sqrt{\frac{m}{2V_0}}|x_1|\,,
            \end{cases}\label{eq:LL_3}
        \end{align}
        with the corresponding classical action 
        \begin{align}
            S &= \sqrt{2m V_0} (|x_0|+|x_1|)-V_0T\,.
        \end{align}    

    \end{itemize}
    See fig.\ \ref{fig:classicalPaths} for an illustration of the three classical paths (blue, yellow and green, respectively).

    \item When the particle traverses the step, with $x_0 < 0 < x_1$, there exists a single classical solution
    \begin{align}
        x(t) = \begin{cases}
            x_0 + v_l t & \text{ when } t \leq \frac{|x_0|}{v_l} \,,\\
            v_r t & \text{ when } t \geq  \frac{|x_0|}{v_l}\,,
        \end{cases}\label{eq:LR}
    \end{align}
    with the velocity on the left of the step $v_l = \sqrt{\frac{2E}{m}}$ and the velocity on the right of the step $v_r = \sqrt{\frac{2(E - V_0)}{m}}$ by the conservation of energy, where the propagation time
    \begin{align}
        T = \frac{|x_0|}{v_l} + \frac{|x_1|} {v_r}
        = \sqrt{\frac{m}{2E}} |x_0| + \sqrt{\frac{m}{2(E-V_0)}} |x_1|\label{eq:LR_T}
    \end{align}
    fixes the energy $E$. The associated classical action assumes the form
    \begin{align}
        S
        =& \sqrt{\frac{mE}{2}} |x_0| + \sqrt{\frac{m(E - V_0)}{2}} |x_1| 
        - V_0 |x_1| \sqrt{\frac{m}{2(E - V_0)}}\,,
    \end{align}
    where $E$ is fixed by the propagation time using equation \eqref{eq:LR_T}. The case $x_1 < 0 < x_0$ follows from the reversibility of classical paths, \textit{i.e.} $x_0 \leftrightarrow x_1$.

    \item Finally, when the boundary conditions lie on the right side of the step $0 < x_0,x_1$, there exists a single classical solution 
    \begin{align}
        x(t) = x_0 + t \frac{x_1-x_0}{T}\,,\label{eq:RR}
    \end{align}
    corresponding to a free particle. Classically, a particle approaching the step from the right only experiences the step at the discontinuity. The particle never reflects the step. The associated classical action assumes the form
    \begin{align}
        S_C = \frac{m(x_1-x_0)^2}{2 T} - V_0 T\,.
    \end{align}
    \end{itemize}
\end{widetext}

%%%%%%%%%%%%%%%%
\subsection{Caustics}
The real classical paths lead to a caustic pattern bounding the interference region in the path integral. As we move from a configuration $(x_0,x_1,T)$ with three real classical paths to one with a single real classical path, two of the classical paths coalesce in a caustic. These caustics correspond to a divergence of the Van Vleck-Morette determinant
\begin{align}
    \frac{\partial^2 S}{\partial x_0 \partial x_1}\,,
\end{align}
where we vary the action of a classical path satisfying the boundary value problem \eqref{eq:BVP} with respect to the boundary values. We evaluate the caustics of the boundary value problem by solving the initial value problem 
\begin{align}
    m \ddot{x} = -V'(x)\,, \quad x(0)=x_0\,, \quad \dot{x}(0) = v_0\,,\label{eq:IVP}
\end{align}
as each pair $(x_0,v_0)$ corresponds uniquely to a classical path. We identify the critical curve 
\begin{align}
    \mathcal{C} = \left\{ (x_0,v_0, T)\,\bigg|\, \frac{\partial x(T)}{\partial v_0} = 0\right\}\,,
\end{align}
in the space of initial conditions using the identity 
\begin{align}
   - \frac{\partial^2 S}{\partial x_0 \partial x_1} = \frac{\partial p_0}{\partial x_1}  = \left[\frac{\partial x_1}{\partial p_0} \right]^{-1}
\end{align}
following from the Hamilton-Jacobi equation $p_0 = - \partial S/\partial x_0$ with $p_0 = m v_0$. The initial value problem maps the critical curve to the caustic curve 
\begin{align}
    x(\mathcal{C}) = \{ (x_0,x_1, T) \,|\, x_1 = x(T) \text{ for some } (x_0,v_0) \in \mathcal{C}\}\,,
\end{align}
in the configuration space of the real-time path integral bounding the interference pattern of the real-time path integral (see the red curves in fig.\ \ref{fig:Tanh_Propagator}). For the step potential, the caustic curve consists of a loop bounding the region with three real classical paths. In the propagator, we can observe the presence of the three solutions in the interference pattern. As one continuously approaches a fold caustic (a differentiable point on the caustic curve), two of the three real classical paths merge. When we approach one of the cusp points (the two non-differentiable points on the caustic curve), the three classical paths coalesce. Outside the caustic region, there exists a unique real solution of the boundary value problem.

\bigskip
For the Heaviside step potential, on the left of the step, there exist three real classical paths in the region
\begin{align}
    x_0,x_1 < 0 \text{ and } |x_0 + x_1| < \sqrt{\frac{V_0}{2 m}}T  \,.
\end{align}
The caustic curve forms a triangle (see the red curves in fig.\ \ref{fig:Step_Propagator}) spanned by the points
\begin{align}
    (x_0,x_1) =(0,0)\,, \left(0, -\sqrt{\frac{V_0}{2m}}T\right)\,, \left(-\sqrt{\frac{V_0}{2m}}T, 0\right)\,.
\end{align}
As we cross the caustic curve, two of the three classical paths merge. The lower left caustic corresponds to the merger of the two bouncing paths. The upper right caustic corresponds to the merger of the direct path and the bounce solution with minimal kinetic energy.

\begin{figure*}
    \centering
    \begin{subfigure}[b]{0.32\linewidth}
        \includegraphics[width=\textwidth]{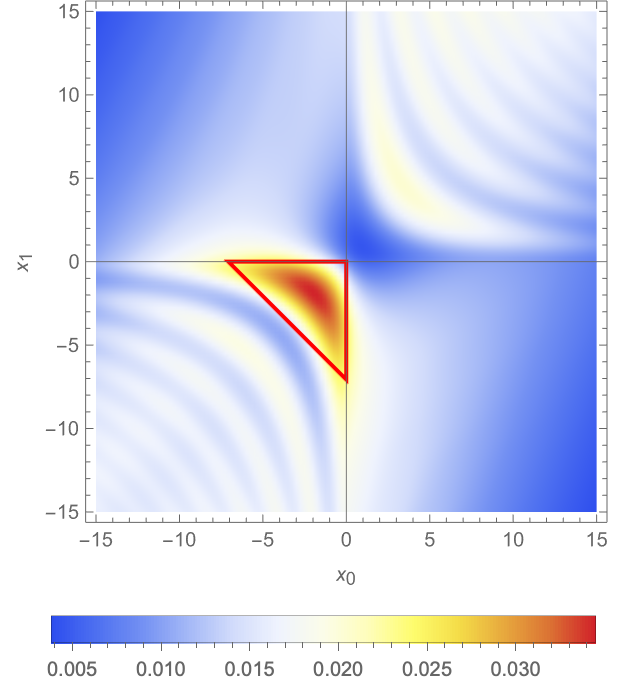}
        \caption{$V_0=0.25$}
    \end{subfigure}
    \hfill
    \begin{subfigure}[b]{0.32\linewidth}
        \includegraphics[width=\textwidth]{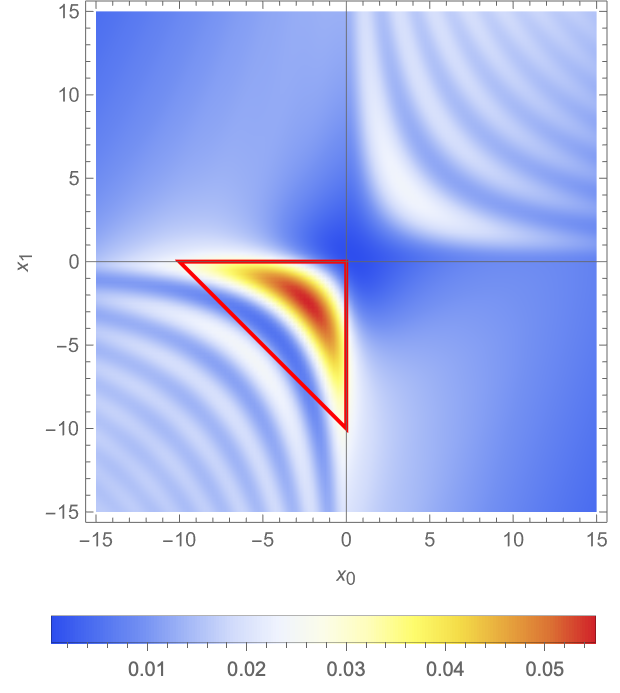}
        \caption{$V_0=0.5$}
    \end{subfigure}
    \hfill
    \begin{subfigure}[b]{0.32\linewidth}
        \includegraphics[width=\textwidth]{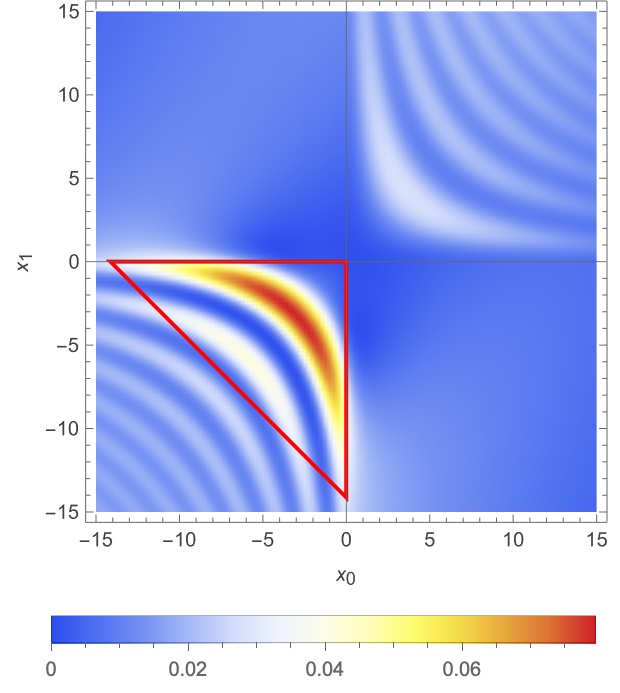}
        \caption{$V_0=1$}
    \end{subfigure}
    \caption{The variation of the real-time propagator of the Heaviside step on the height of the step $V_0=0.25,0.5,$ and $1$ (left to right) for $T=10, m=1$. The red lines mark the caustics.}\label{fig:step_V0}
% \end{figure}

% \begin{figure}
    \centering
    \begin{subfigure}[b]{0.32\linewidth}
        \includegraphics[width=\textwidth]{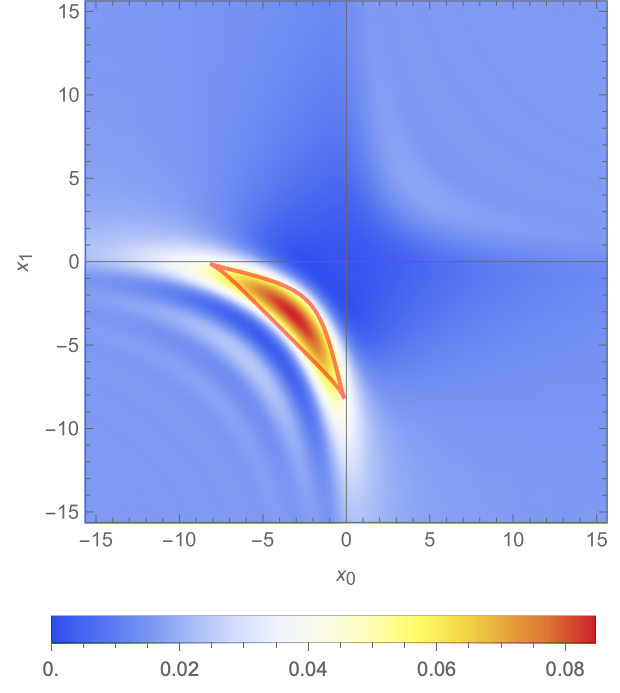}
        \caption{$V_0=1$}
    \end{subfigure}
    \hfill
    \begin{subfigure}[b]{0.32\linewidth}
        \includegraphics[width=\textwidth]{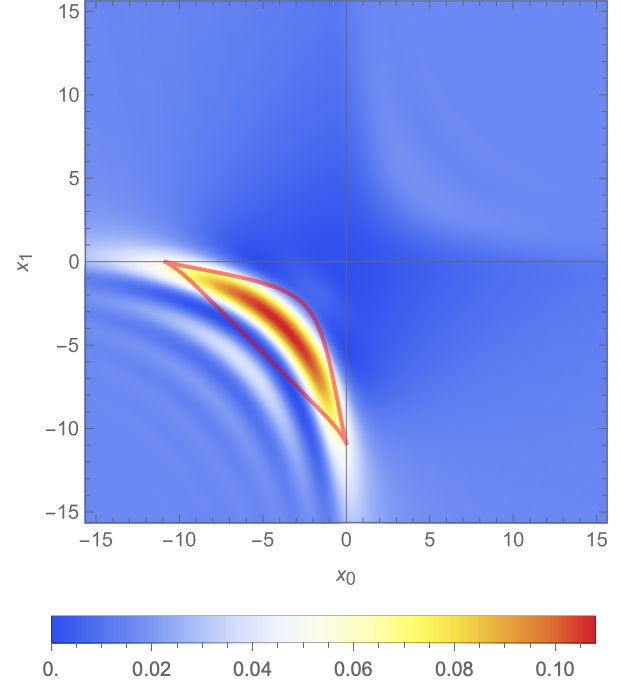}
        \caption{$V_0=1.5$}
    \end{subfigure}
    \hfill
    \begin{subfigure}[b]{0.32\linewidth}
        \includegraphics[width=\textwidth]{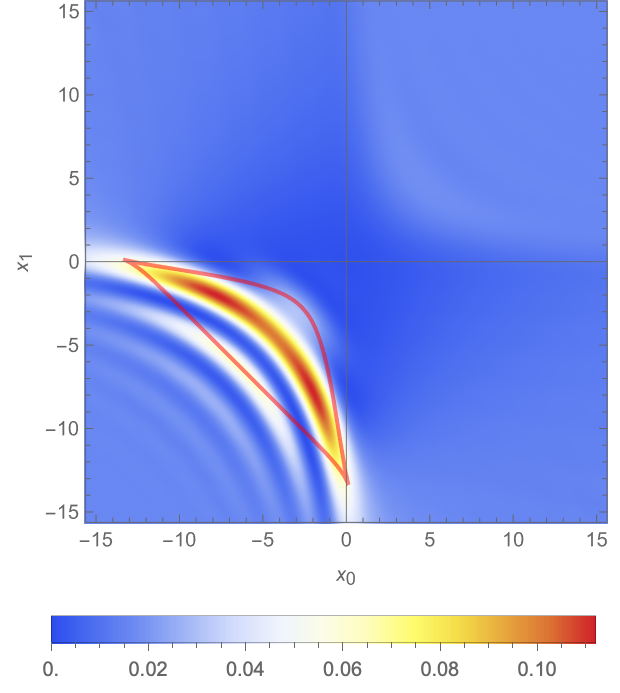}
        \caption{$V_0=2$}
    \end{subfigure}
    \caption{The variation of the real-time propagator of the smooth step on the height of the step $V_0=1,1.5,$ and $2$ (left to right) for $T=10, m=1,$ and $\alpha=1$. The red curves mark the caustics.}\label{fig:smooth_V0}
\end{figure*}

From the caustics of the Heaviside step potential, we observe a strong dependence on the height of the step. This is reflected in the interference pattern in the path integral (see fig.\ \ref{fig:step_V0}). Similar behaviour is observed for the smooth step function (see fig.\ \ref{fig:smooth_V0}). As the step becomes higher, the region with three real classical paths extends. In \cite{Feldbrugge:2023a}, we showed that the caustics of the classical theory provide a coarse description of the nature of wave packets in the quantum theory. Consequently, when starting with a sharply localised initial state on the left of the step, we can measure the height of the step from the reflected wave packet on the left of the step at a later time. This is unlike the classical theory, where a bouncing particle in a single experiment does not include any information about the height of the step.

%%%%%%%%%%%%%%%%%%%%%%%%%%%%%%%
\subsection{Complex semi-classical paths}\label{sec:complex-paths}
The Wentzel–Kramers–Brillouin (WKB) approximation approximates the real-time path integral in terms of the classical paths,
\begin{align}
    G(x_1,x_0;T) \approx \Theta(T) \sqrt{\frac{i}{2 \pi \hbar}} \sum_{x_C} \sqrt{\frac{\partial^2 S}{\partial x_0 \partial x_1}} e^{i S[x_C]/\hbar}\,,
\end{align}
where the sum runs over the relevant classical paths solving the boundary value problem \cite{Wentzel:1926, Kramers:1926, Brillouin:1926}. Formally, based on Picard-Lefschetz theory, a classical path should be included in the WKB approximation if and only if the steepest ascent manifold of the classical path intersects the space of real-valued paths interpolating between $x_0$ and $x_1$ (see for details \cite{Feldbrugge:2023c, Feldbrugge:2023a}). In practice, these intersections are difficult to evaluate numerically.

\begin{figure*}
    \centering
    \begin{subfigure}[b]{0.49\linewidth}
        \includegraphics[width=\textwidth]{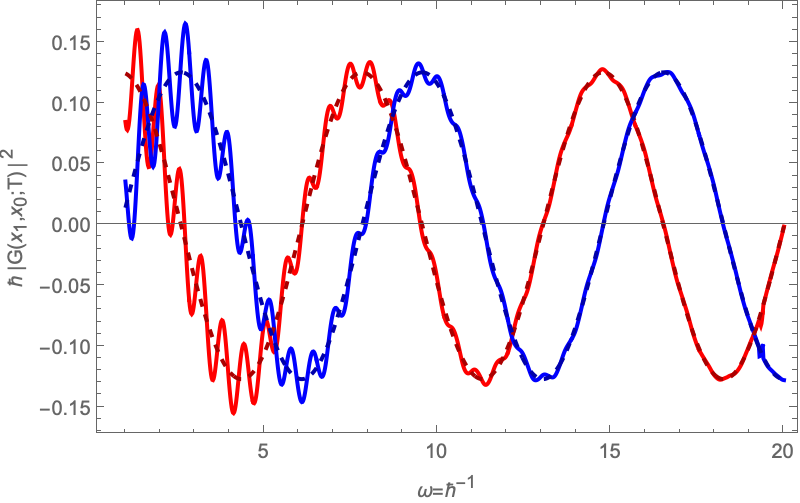}
        \caption{The real classical path. \phantom{test test test} \phantom{test test test} \phantom{test test test test} }
    \end{subfigure}
    \begin{subfigure}[b]{0.49\linewidth}
        \includegraphics[width=\textwidth]{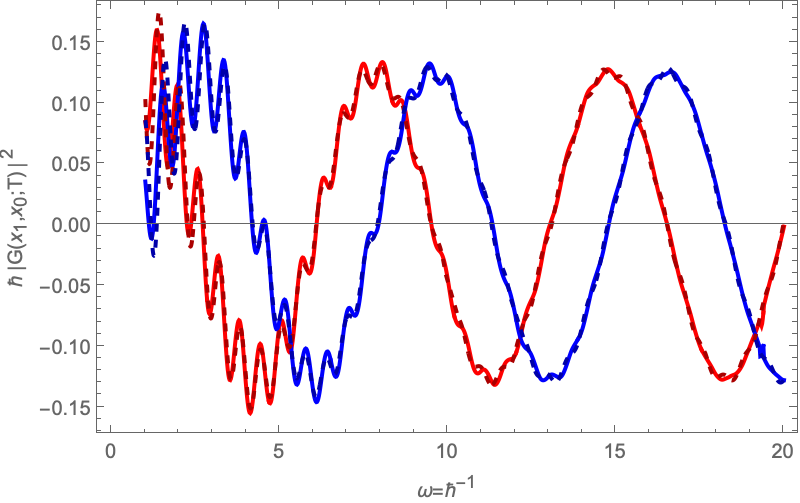}
        \caption{The real classical path and the complex semi-classical path associated with the caustic}
    \end{subfigure}
    \hfill
    \caption{The real (red) and imaginary parts (blue) of the real-time propagator  $G(x_1,x_0;T)$  (solid) and the WKB approximation including (left) only the real classical paths and (right) the real and the caustic complex semi-classical path for the smooth step with the boundary conditions $x_0=-5, x_1=-9.25$ with the propagation time $T=10$ and the strength of the step $V_0=1$ and the steepness parameter $ \alpha = 5$. }\label{fig:LEFT}
\end{figure*}

Let's consider the path integral for a particle reflecting off the step (with $x_0,x_1 <0$ outside the caustic region). The WKB term associated with the real classical path provides a good first approximation of the real-time propagator (see the left panel of fig.\ \ref{fig:LEFT}). As we will demonstrate, the remaining high-frequency oscillations, missed by the real classical path, are the consequence of a complex solution to the boundary value problem. 

\begin{figure*}
    \centering
    \begin{subfigure}[b]{0.49\linewidth}
        \includegraphics[width=\linewidth]{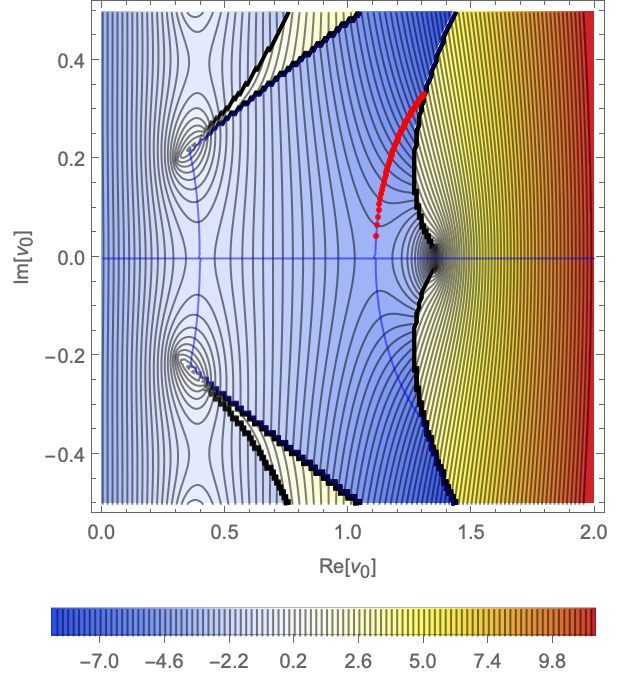}
        \caption{The complex semi-classical path in the complex initial velocity plane for the initial position $x_0=-4$ and the final position $x_1=-6.75, -6.7, \dots,-3.95$.
        \phantom{test test test test test test}
        \phantom{test test test test test test}
        \phantom{test test test test test test}
        \phantom{test test test test test test}}
    \end{subfigure}
    \begin{subfigure}[b]{0.49\linewidth}
        \includegraphics[width=\linewidth]{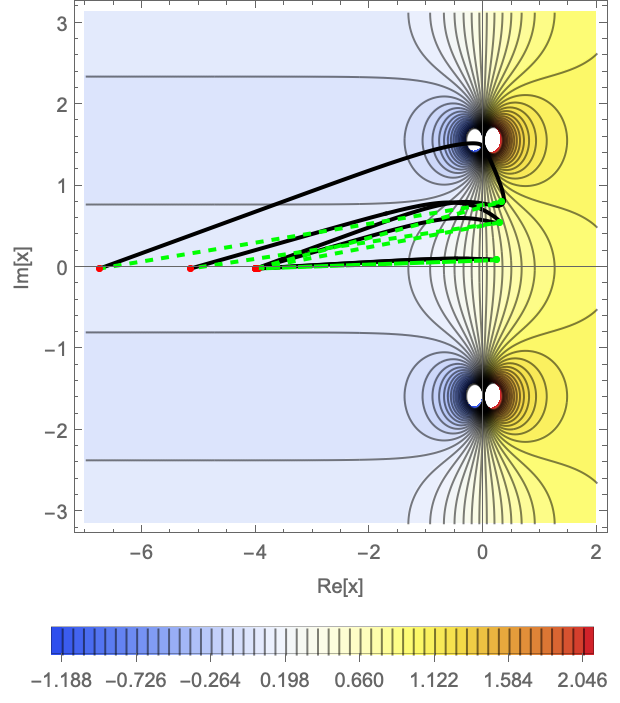}
        \caption{The complex semi-classical path (black) for the initial position $x_0=-4$ and the final position for $x_1=-3.95, -5.15, -6.75$ approaching the singularity of the potential $V(x)$ at $x=i \pi/2$. The green line shows the corresponding contour running from the initial position $x_0$ to the final position $x_1$ via the turning point $x_t$.}\label{fig:10b}
    \end{subfigure}
    \caption{The evolution of a complex semi-classical path as we continuously deform the final position $x_1$ of the boundary value problem, for $m=1, V_0=1, T=10$ and $\alpha=1$.}\label{fig:Complex_paths}
\end{figure*}

Starting from inside the caustic region -- with three real solutions to the boundary value problem -- and continuously moving in configuration space to the left, two of the real classical paths coalesce as we cross the caustic. Outside the caustic region, these classical paths turn into a complex solution of the boundary value problem (see fig.\ \ref{fig:Complex_paths} for one of these complex solutions). The semi-classical paths form complex conjugate pairs, one of which remains relevant while the other complex path becomes irrelevant to the path integral. When including the saddle point approximation of this complex semi-classical path, we recover the high-frequency mode in the propagator (see the right panel of fig.\ \ref{fig:LEFT}). Generally, the complex semi-classical path for which the real part of the exponent $i S[x_C]$ is negative contributes to the path integral after crossing the caustic.

However, as we continue our journey outside the caustic region, we find that, at some point, the complex semi-classical path intersects one of the singularities of the analytic continuation of the potential located along the imaginary axes 
\begin{align}
    x_s = \frac{i \pi}{\alpha}\left(n+\frac{1}{2}\right)\,.
\end{align}
See fig.\ \ref{fig:Complex_paths} for an illustration of this phenomenon. After this \textit{singularity crossing}, the semi-classical path ceases to exist as a naive solution to the boundary value problem. When investigating the evolution of the classical action, we find that the action evolves smoothly and encounters a branch-cut when the semi-classical path crosses the singularity. In the limit of the Heaviside step, $\alpha \to \infty$, the singularities approach the real line and the singularity crossing occurs instantly when two real classical paths merge in a caustic.

We observed this phenomenon for the first time while studying tunnelling in a Rosen-Morse barrier, and proposed to analytically continue the complex semi-classical paths beyond the singularity crossing to recover the WKB approximation \cite{Feldbrugge:2023a,Feldbrugge:2023b}. In fact, we showed that these analytic continuations are necessary to model quantum tunnelling in terms of semi-classical paths. Recently, we realised that this is a general phenomenon which occurs when the classical equation of motion is not of Painlevé type. At the start of the 20th century, Painlevé realised that the second-order ordinary differential equations whose solutions are of Painlevé type -- the only movable singularities are simple poles -- can be reduced to $50$ canonical forms \cite{Painleve:1900, Painleve:1902, Gambier:1910}. Out of these $50$ differential equations, six led to the introduction of new special functions, currently known as the Painlevé transcendents. As the Painlevé property is rare in non-linear second-order differential equations, solutions of the equation of motion of general physical systems develop branch-cuts that will generally be crossed by semi-classical paths.

% \begin{figure*}
%     \centering
%     \begin{subfigure}[b]{0.49\linewidth}
%         \includegraphics[width=\linewidth]{complex_v0.png}
%         \caption{The complex semi-classical path in the complex initial velocity plane for the initial position $x_0=-4$ and the final position $x_1=-1.3, -1.2, \dots,0.3$.}
%     \end{subfigure}
%     \begin{subfigure}[b]{0.49\linewidth}
%         \includegraphics[width=\linewidth]{SingularityCrossing.png}
%         \caption{The complex semi-classical path for the initial position $x_0=-4$ and the final position $x_1=-1.3, -1.1, \dots,0.3$ approaching the singularity of the potential $V(x)$ at $x=\pi/2$.}
%     \end{subfigure}
%     \caption{The evolution of a complex semi-classical path as we continuously deform the final position $x_1$ of the boundary value problem, for $m=1, V_0=1$ and $\alpha=1$.}\label{fig:Complex_paths}
% \end{figure*}

To analytically continue the semi-classical path beyond a singularity crossing, we note that as the action is an integral over time, 
\begin{align}
    S[x] = \int_0^T L(x,\dot{x})\mathrm{d}t\,.
\end{align}
By deforming the integration domain $[0,T]$ into the complex plane while keeping the endpoints fixed, we obtain an equivalence class of semi-classical paths consisting of solutions to the boundary value problem defined along complex time-domain contours starting at $0$ and terminating at $T$,
\begin{align}
    [x]_C = \bigg\{(&x(\lambda), t(\lambda)) \,\bigg|\, 
    \frac{t' x'' - x' t''}{(t') ^3} = - \nabla_x V(x)\,,\nonumber\\
    &t(0)=0\,, t(1) = T\,, x(0)=x_0\,, x(1)=x_1 \bigg\}\,,
\end{align}
where the time parameter $t(\lambda)$ of the path $(x(\lambda), t(\lambda))$ for $0 \leq \lambda \leq 1$ is a complex contour that can be continuously deformed to the interval $[0,T]$ without crossing singularities. In this definition, the primes denote differentiation with respect to the affine parameter $\lambda$. By construction, the equivalence class of the semi-classical paths corresponds to a single classical action (it is the defining invariant). In equation \eqref{eq:Action}, we eliminated the time domain in favour of the path. For real classical paths, the extension of classical paths to these equivalence classes does not add to the theory. However, at a singularity crossing, the naive semi-classical path ceases to exist, but the equivalence class remains well-defined with a meaningful contribution to the WKB approximation.

\begin{figure}
    \centering
    \begin{tikzpicture}
        \node[anchor= south west, inner sep=0] (img) at (0,0) {\includegraphics[width=\linewidth]{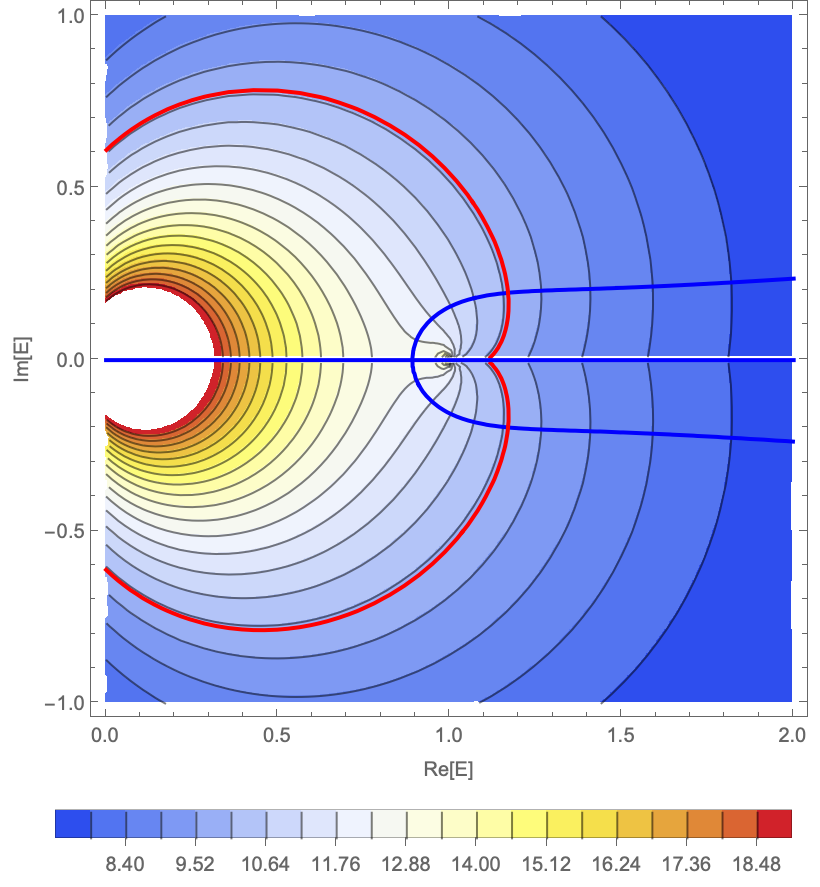}};
        \filldraw[white] (5.35,4.85) circle (0.1);
        \draw[thick, green] (5.35,4.85) circle (0.1);
        \filldraw[yellow] (5.2,5.575) circle (0.1);
        \filldraw[green] (5.35,6.3) circle (0.1);
        \node[above right] at (5.35,6.3) {\textbf{caustic saddle}};
        \node[above right] at (5.25,5.47) {\textbf{topological saddle}};
        % \node[above right] at (5.35,4.85) {\textbf{caustic saddle}};
    \end{tikzpicture}
    \caption{The propagation time $T$ in the complex energy $E$ plane, for $m=1, \alpha=5, x_0=-5,x_1=-9.25$. The background represents the real part of the propagation time. The blue lines mark energies for which the propagation time is real valued. The red curve represents the energies for which the real part of the propagation time is $T=10$.}\label{fig:T_analytic}
\end{figure}

To find the complex semi-classical paths beyond the singularity crossing, we solve the propagation time for the bounce solution \eqref{eq:t2} for complex energies $E$. The complex semi-classical path associated with the fold caustic is given by the intersection of the red and blue curves in the upper half plane (see fig.\ \ref{fig:T_analytic}). Given the energy, we can evaluate the associated classical action \eqref{eq:S2} and contribution to the WKB approximation. In the left panel of fig.\, \ref{fig:LEFT} we observe that this semi-classical path, beyond the singularities crossing, is indeed responsible for the high frequency mode in the propagator. The quantum reflection on the left of the step potential is governed by the semi-classical path associated with the caustic.

% Irrelevant saddle point open circle

One of the complex semi-classical paths emerging at a caustic tends to be relevant to the path integral in the vicinity of the caustic. However, as we move away from the caustic, the semi-classical path can undergo a Stokes phenomenon, after which its contribution to the path integral switches off. This happens when the steepest ascent manifold of the complex semi-classical path intersects another classical path (for details see \cite{Feldbrugge:2023d, Feldbrugge:2023a}). The Stokes lines generally emanate from the cusp caustics and form wedges in parameter space. At the Stokes lines, the real parts of two classical paths, $x_{C_1}$ and $x_{C_2}$, coincide parameter space, \textit{i.e.},
\begin{align}
    \text{Re}(S[x_{C_1}]) = \text{Re}(S[x_{C_2}])\,.
\end{align}
For the step potential, the Stokes lines form two wedges emanating from the two cusp points (see the green curves in fig. \ref{fig:Tanh_Propagator}). In the upper right and lower left regions of parameter space, the semi-classical path associated with the caustic is relevant to the path integral. In the upper left and lower right regions, the semi-classical path does not contribute. For a more detailed description of Stokes phenomena in the real-time path internal, we refer to \cite{Feldbrugge:2023a}. When approaching the Heaviside step, the Stokes lines turn into straight lines along the horizontal and vertical axes of the parameter space (see the green lines in fig.\ \ref{fig:Step_Propagator}).

Fig.\, \ref{fig:10b} shows three complex paths emanating from the caustic.  The black curves are trajectories for real time.  The dashed lines are straight line segments to and from the turning point. The action is identical along the two paths, as are the endpoint values of time and position.  The real-time trajectories navigate a finite distance around the turning point, which corresponds to a branch cut: already on the real line, $t\propto \pm \sqrt{x-x_t}$, with a sign change at the turning point $x=x_t$.

As one moves the end point $x_1$ further to the left at a fixed time interval, the energy will increase, and one might expect a sufficiently high energy/velocity particle to pass over the step and continue to infinity rather than reflect off the step.  Numerical inspection shows that the energy stays complex even as $x_1$ goes to negative infinity, leaving the puzzling question of what happens to a particle encountering the step with a real energy above the step height: for a steep step (or large $\alpha$), some fraction reflects.

Here we note that the action is a physical quantity, and while a given (complex) energy maps to a specific action through eqn.\, \eqref{eq:sx}, the converse does not have to be one-to-one.  We discuss this further in the next subsection.

\subsection{Quantum Reflection: Topological Action}

The complex trajectories in the previous subsection correspond to complex energies.  Yet, positive real energy wave packets certainly partially reflect on the potential, and we can examine real energy complex trajectories.  

\begin{figure*}
    \centering
    \includegraphics[width=0.7 \textwidth]{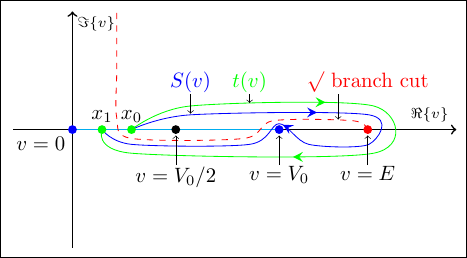}
    \caption{
      Saddles for a reflection of a particle from the left.  We can choose the contour along the real line, with judicious choices at the regular pole at $v=V_0$, and the standard $dt/dv\propto 1/\sqrt{v}$ branch turn at the turning point at $v_t=E$. For the action, the square root branch cut $dS/dv\propto \sqrt{v}$, with $S\propto t^3$ at the turning point, resulting in a different path through the pole.  The time interval is real, while the action acquires a complex residual.
    }
    \label{fig:sleft}
% \end{figure}

% \begin{figure}
    \centering
    \includegraphics[width=0.7 \textwidth]{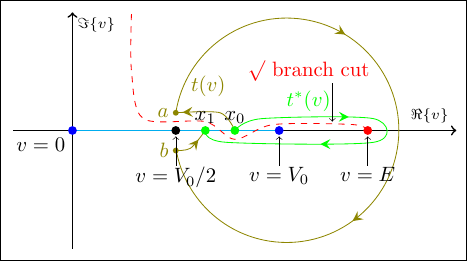}
    \caption{
      Saddles for a reflection of a particle from the right.  A direct contour choice of tracing the real line to the turning point and back corresponds to a counterintuitive notion of a particle running to infinity and back, instead of reflecting at the step. Instead, the red saddle illustrates a contour from the same equivalence class, \textit{i.e.,} the same action, that moves consistently with intuition, first to the step, and then looping 
    }
    \label{fig:sright}
\end{figure*}

For $E>V_0$, a particle from the left will partially reflect, and similarly, a particle from the right will also partially reflect. To construct real energy reflecting saddles, transform to the $v=V(x)$ coordinate.  The turning point maps to $v=E$. For $E>v_0$ it might appear as `business as usual', a particle goes to the turning point along the real $v$ line, turns around the $t=0$ branch cut at $v_t=E$, and comes back along the real $v$ line.  One might puzzle that this saddle always appears to reflect on the far right, while we expect the reflection to occur near $v=V_0/2$.  At first sight, it appears even more unusual that for initial positions on the right side of the barrier, the ``real'' trajectory doesn't even appear to move towards the step, as sketched by the green curve in fig.\, \ref{fig:sright}. 

A more careful examination elucidates these subtleties.  The time variable appears complex, though this can be cancelled with the appropriate integration constant $C$.  It exhibits a log singularity at $v=V_0$, and gains a complex constant past the singularity.  On the return, the square root branch cut changes the time constant at $v<V_0$.  By passing just above the singularity on the outbound trajectory and just below the singularity on the return, the complex phase conveniently cancels, and the total elapsed time $T=t_0+t_1$ is real.  

We can express the action in terms of only the kinetic term
\begin{align}
S_k&=-ET + \int  m\dot{x}^2 dt\\
& =
\frac{\sqrt{E-1}}{\sqrt{2}\alpha}\tanh^{-1}(\sqrt{z_+})
-\frac{\sqrt{E}}{\sqrt{2}\alpha}\tanh^{-1}(\sqrt{z_-}) \\
&= S_++S_- - ET
\end{align}

For $E>V_0$ this arctanh in $S_+$ acquires a $\pi i$ complex phase, $Im(S_+)=\sqrt{E-V_0} \pi/\sqrt{2}\alpha=iS_I/2$.  This complex component accounts for the instanton damping of the reflection coefficient $S_I$.  In the limit $\alpha^2>2\pi(E-V_0)$, the damping becomes small, reverting to the step function behaviour. 

The left saddles are illustrated in fig.\ \ref{fig:sleft}. The action contour picks up an imaginary component from the log singularity, while the time contour does not.  This can be seen from the topology that the kinetic action term is the cube of time near the turning point.  As the contour in $v$ turns to backtrack, $t$ changes by a minus sign due to the square root branch cut, and passes on the opposite side of the log singularity on the return path to cancel the imaginary contribution.  If we think of the sign change as a $\pi$ rotation, the action rotates by $3\pi$ at the turning point, crossing the log singularity once in the process, and picking up $2\pi i$. 

For $x_0,x_1>0$, two equivalent contours are illustrated in fig.\, \ref{fig:sright}.  The green contour is the direct real axis saddle extended from the previous discussion.  It appears rather non-intuitive, since the reflection is occurring at the step, while the contour doesn't go anywhere near the step.  An alternate contour $C$ with the same outcome is shown in brown.  We decompose $C=C_-+C_o +C_+$, with $C_-$ going from $x_0$ to $a$; $C_o$ goes from $a$ to $b$ looping around the branch cuts, with $C_+$ going from $b$ to $x_1$. The time variable $t$ is real and increasing along $C_-$ and $C_+$, going from $t_0$ at $x_0$ to 0 $a$, then taking a complex loop from $a$ to $b$ along $C_o$, and finally going along the real axis from $b$ to $t_1$ along $C_+$.   

From eqn.\, \eqref{eqn:tx}, we choose the integration constant $c_t$ to be purely imaginary.  The real axis reflection points satisfy $Re[t(a)]=0$, for which we choose $c_t$ such that $t(a)=0$, implementing the interpretation that $t$ is taking an imaginary trajectory along $C_0$, with $t(x_t)=c_t$.  

\begin{figure}
    \centering
    \includegraphics[width=\linewidth]{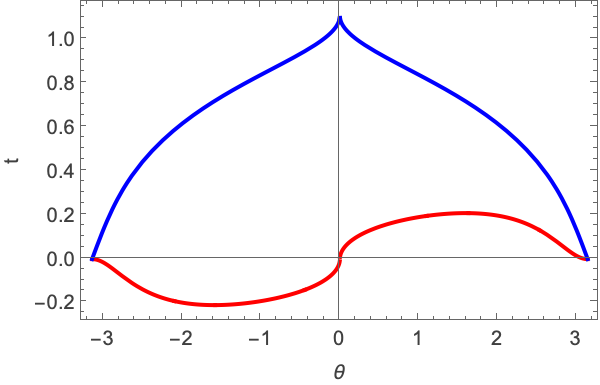}
    \caption{Time variable $t$ along the instanton saddle $C_0$ for $E=2$, with red the real and blue the imaginary part of $t$ as a function of $\theta$.}\label{fig:c0}
\end{figure}

For $E=2,\ \alpha=1$ we find $a=b\sim -0.26$. A convenient
parameterization for $C_0$ is $v=(E-a) e^{i\theta}/2+(E+a)/2$, with
$-\pi<\theta<\pi$.  Along $C_0$, $Im(t)$ reaches a maximum at
$\theta=0$.  This segment is shown in fig.\, \ref{fig:c0}.

The action is similarly real along $C_-,C_+$, and complex along $C_o$, which corresponds to the instanton.  The matching points have $a=b\sim 0$ near zero (slightly negative).  Since a branch cut runs between $a$ and $b$, we can shift $a$ just above the real line, and $b$ just below, as illustrated in fig.\, \ref{fig:sright}.

The action is similarly decomposed, with the real part arising from $C_+,C_-$, and the imaginary part from $C_0$.  As noted, the action contour passes above the $\tanh$ branch cut, leading $S_0$ to be purely imaginary.

In the limit of large $\alpha$, the imaginary part of the action is suppressed by $1/\alpha$, with no imaginary part in the step function limit.  This recovers the standard result, with an explicit saddle construction and a real-time propagator.

In the step function limit, the action of this reflecting saddle is real, and its contour runs along the real $v$ axis.  While one generally associates this property with classical saddles, this construction elucidates the range of semi-classical and quantum saddles.

We call these heuristically derived actions {\it topological}, to contrast with the {\it caustic} actions in the previous section. Numerically, the topological actions have a minimum energy $E>V_0$, thus near the caustic, these {\it topological} actions do not exist. Numerically, we observe that the topological action and the caustic actions have similar values, and converge rapidly to each other as the potential steepens (\textit{i.e.} for large $\alpha$, or large $x_0,\ x_1$). This corresponds to multiple energy solutions for a given action $S(E_i)$. Since the propagator depends only on the action, the topological action provides a simple semi-analytic instanton with real energy.  We also note that the topological action starts real at $E=V_0$, and its imaginary component approaches the saddle point action as energy increases.

We will numerically study their contributions in regions where both exist in the next section.

%%%%%%%%%%%%%%%%%%%%%%%%%%%%%%%
\section{Fourier and Laplace analyses}\label{sec:fourier}
In the previous section, we evaluated the Feynman path integral and analysed its behaviour in terms of the classical and complex semi-classical paths. It follows from Picard-Lefschetz theory that a complex semi-classical path is relevant to the path integral if and only if its steepest ascent manifold intersects the original integration domain. However, in practice, this formal condition is hard to implement for high-dimensional integrals. The geometry of the caustics and Stokes lines is very helpful in suggesting the relevance of the different classical paths. However, this analysis may still miss the contribution of a complex semi-classical path. 

In this section, we develop a systematic method to infer the relevant real and complex (semi-)classical paths directly from the Feynman path integral, evaluated either with spectral methods or numerical evaluation methods \cite{Feldbrugge:2025}. Concretely, we consider the Fourier and Laplace transform of the propagator with respect to the reciprocal of the Planck constant $\omega = \hbar^{-1}$. The Fourier transform peaks at the real part of the relevant classical actions. Irrelevant instantons do not show up in the Fourier analysis. The Laplace transform enables us to infer the imaginary parts of the relevant classical paths.

\subsection{The Fourier transform}
In the semi-classical limit, the path integral approaches the WKB approximation, consisting of a contribution of each relevant classical path $x_C$,
\begin{align}
    G(x_1,x_0;T) \approx \Theta(T) \sqrt{\frac{i}{2 \pi \hbar}}\sum_{x_C} \sqrt{\frac{\partial^2 S[x_C]}{\partial x_0 \partial x_1}} e^{i S[x_C]/\hbar}\,.
\end{align}
Fourier transforming the WKB approximation with respect to the reciprocal of Planck's constant $\omega = \hbar^{-1}$, we construct the spectrum
\begin{align}
    \mathcal{F}(\tau) &= \int_A^B \sqrt{\frac{2 \pi}{i \omega}}G(x_1,x_0;T) e^{i \omega \tau}\mathrm{d}\omega \nonumber\\
    &\approx \Theta(T)
    \sum_{x_C } \sqrt{\frac{\partial^2 S_C}{\partial x_0 \partial x_1}} \int_A^B e^{i \omega S_C} e^{i \omega \tau}\mathrm{d}\omega\\
    &= i \Theta(T)\sum_{x_C } \sqrt{\frac{\partial^2 S_C}{\partial x_0 \partial x_1}} \frac{e^{i A(S_C + \tau)} - e^{i B(S_C + \tau)}}{S_C + \tau}
\end{align}
evaluated over the positive interval  $0<A<B$. The spectrum $\mathcal{F}$ peaks when the denominator $|S_C + \tau|$ is small. Since
\begin{align}
    \left| \frac{1}{S_C + \tau}\right|^2 = \frac{1}{(\text{Re}[S_C]- \tau)^2 + \text{Im}[S_C]^2}
\end{align}
for real $\tau$, the modulus of the spectrum $|\mathcal{F}(\tau)|^2$ peaks at the real part of the classical actions of the relevant classical paths. In principle, we can infer the imaginary part of the classical action from the shape and height of the peak.

\begin{figure*}
    \centering
    \begin{subfigure}[b]{0.32\linewidth}
        \includegraphics[width=\textwidth]{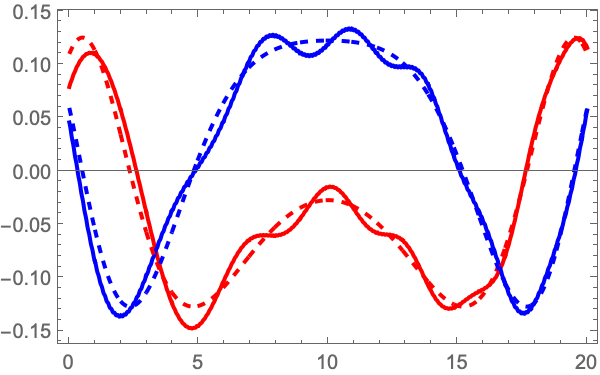}
        \caption{$\hbar=1$}
    \end{subfigure}
    \hfill
    \begin{subfigure}[b]{0.32\linewidth}
        \includegraphics[width=\textwidth]{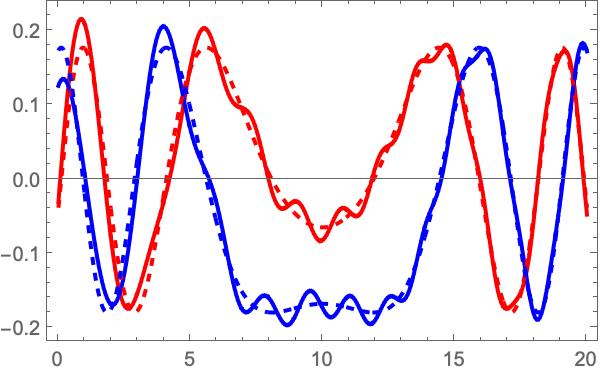}
        \caption{$\hbar=0.5$}
    \end{subfigure}
    \hfill
    \begin{subfigure}[b]{0.32\linewidth}
        \includegraphics[width=\textwidth]{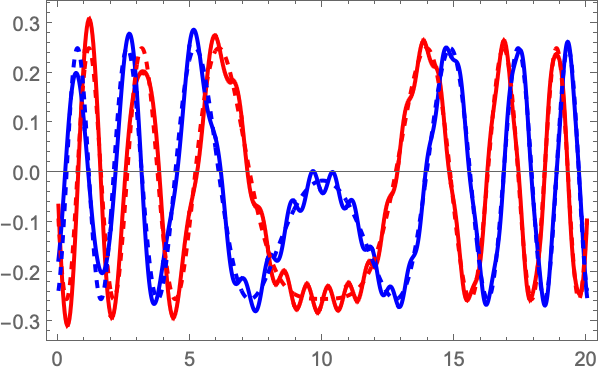}
        \caption{$\hbar=0.25$}
    \end{subfigure}\\
    \begin{subfigure}[b]{0.32\linewidth}
        \includegraphics[width=\textwidth]{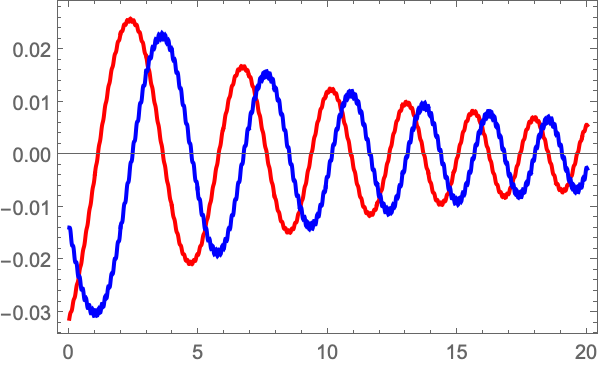}
        \caption{$\hbar=1$}
    \end{subfigure}
    \hfill
    \begin{subfigure}[b]{0.32\linewidth}
        \includegraphics[width=\textwidth]{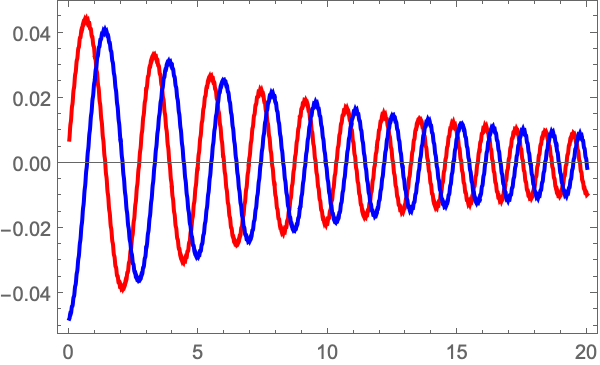}
        \caption{$\hbar=0.5$}
    \end{subfigure}
    \hfill
    \begin{subfigure}[b]{0.32\linewidth}
        \includegraphics[width=\textwidth]{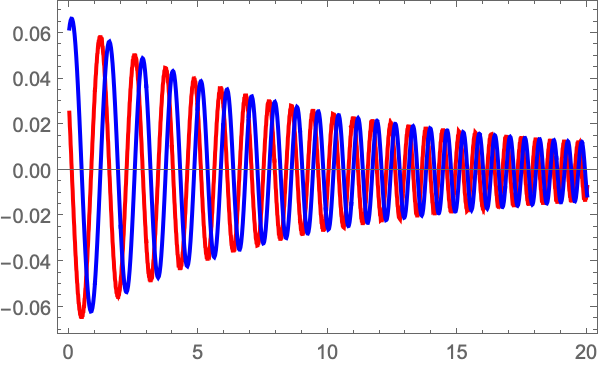}
        \caption{$\hbar=0.25$}
    \end{subfigure}\\
    \caption{Upper panels: The real and imaginary parts of the Heaviside step propagator (solid) and a free particle (dashed) with the initial position $x_0=10$, mass $m=1$, step height $V_0=1$, and propagation time $T=10$ as a function of the final position $x_1$. Lower panels: The difference between the propagator and a free particle. The real and imaginary parts of the propagator are plotted in red and blue.}\label{fig:left_left}
\end{figure*}

\begin{figure*}
    \centering
    \begin{subfigure}[b]{0.49\linewidth}
        \includegraphics[width=\textwidth]{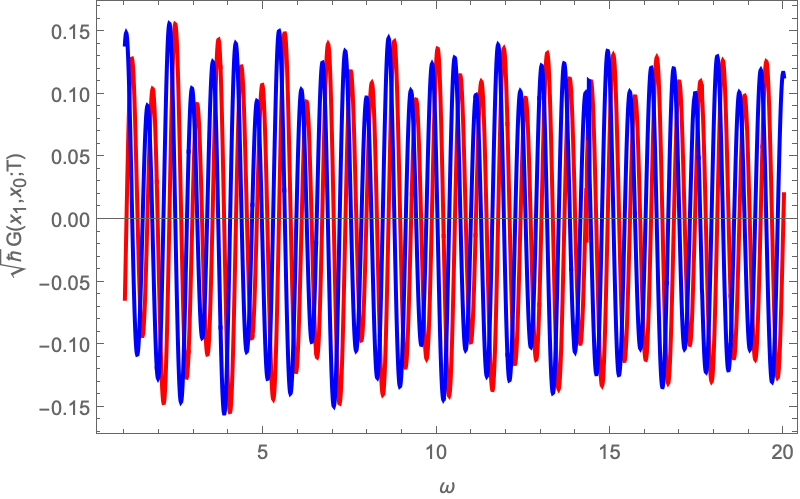}
    \end{subfigure}
    \hfill
    \begin{subfigure}[b]{0.49\linewidth}
        \includegraphics[width=\textwidth]{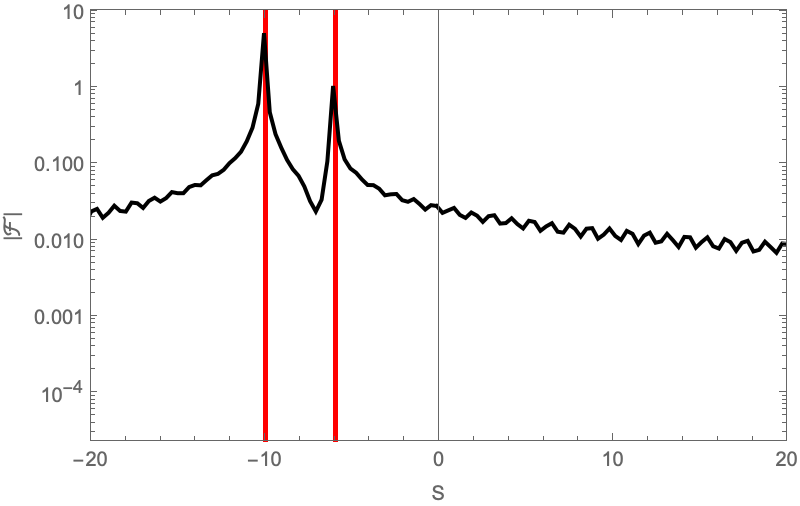}
    \end{subfigure}
    \caption{The path integral of the Heaviside step (left) and corresponding spectrum (right) of a particle travelling from $x_0=5$ to $x_1=4$ with a unit step $V_0=1$. The red lines represent the two classical actions.}\label{fig:RR}
\end{figure*}

We illustrate the Fourier transform for the propagator of the Heaviside step potential with the initial and final boundary conditions on the right of the step. Like we observed on the left of the step, the propagator consists of both a long and a short wavelength mode (see fig.\ \ref{fig:left_left}). The long wavelength mode corresponds to the real classical path, whereas the short wavelength mode corresponds to a quantum reflection. The associated spectrum includes two peaks (see fig.\ \ref{fig:RR}). The left peak matches the action of a path directly propagating between the boundary points, with the action
\begin{align}
    S = \frac{m(x_1-x_0)^2}{2T} - V_0 T\,.
\end{align}
The right peak corresponds to a quantum reflection bouncing of the step, with the action 
\begin{align}
    S = \frac{m(x_1+x_0)^2}{2T} - V_0 T\,.
\end{align}
Note that there does not exist a naive semi-classical path corresponding to this trajectory in the Heaviside step potential, as the particle directly undergoes a singularity crossing as $x_0$ or $x_1$ crosses the step at $x=0$ in the limit $\alpha \to \infty$. The action suggests the presence of the minimal energy bounce.  Remarkably, the action of this semi-classical path is real, matching the observation that the quantum reflection of the Heaviside step is preserved in the semi-classical limit $\hbar \to 0$. 

\begin{figure*}
    \centering
    \begin{subfigure}[b]{0.49\linewidth}
        \includegraphics[width=\textwidth]{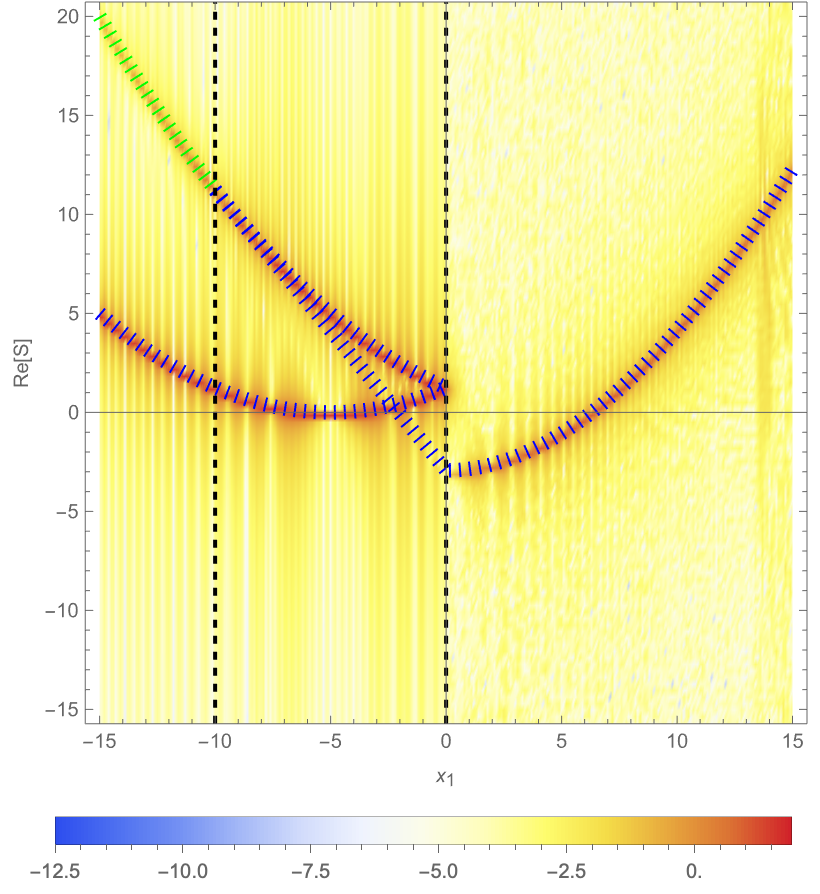}
        \caption{$x_0=-5$}\label{fig:fourier_L}
    \end{subfigure}
    \hfill
    \begin{subfigure}[b]{0.49\linewidth}
        \includegraphics[width=\textwidth]{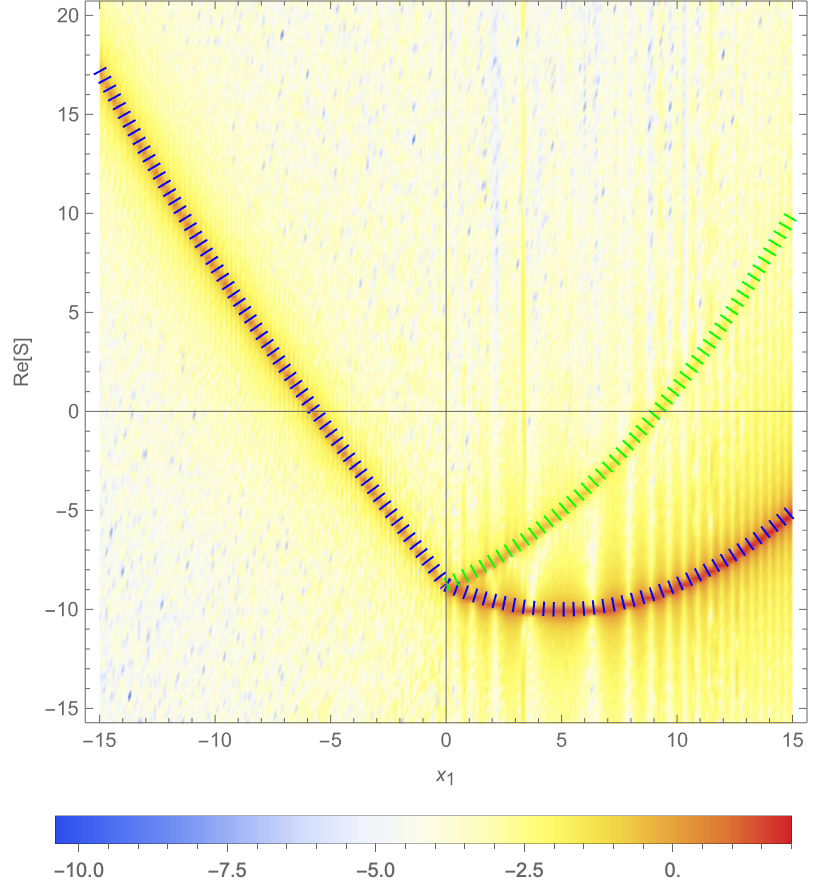}
        \caption{$x_0=5$}\label{fig:fourier_R}
    \end{subfigure}
    \caption{A logarithmic plot of the Fourier transform of the Heaviside step propagator with respect to the frequency parameter $\omega = 1/ \hbar$ for the step height $V_0=1$, mass $m=1$ and propagation time $T=10$. The blue dashed lines correspond to the actions of the real classical paths. The green dashed lines correspond to the real part of the actions of the relevant complex semi-classical paths. The dashed vertical lines mark the caustics.}\label{fig:fourier}
% \end{figure}

% \begin{figure}
    \centering
    \begin{subfigure}[b]{0.49\linewidth}
        \includegraphics[width=\textwidth]{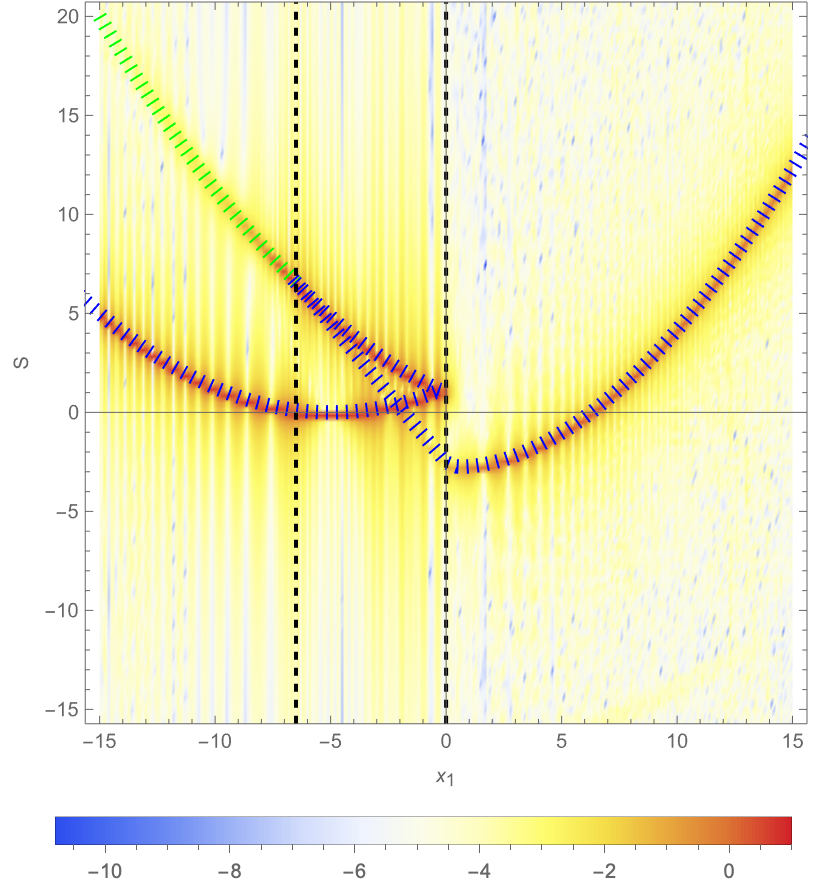}
        \caption{$x_0=-5$}\label{fig:fourier_smooth_L}
    \end{subfigure}
    \hfill
    \begin{subfigure}[b]{0.49\linewidth}
        \includegraphics[width=\textwidth]{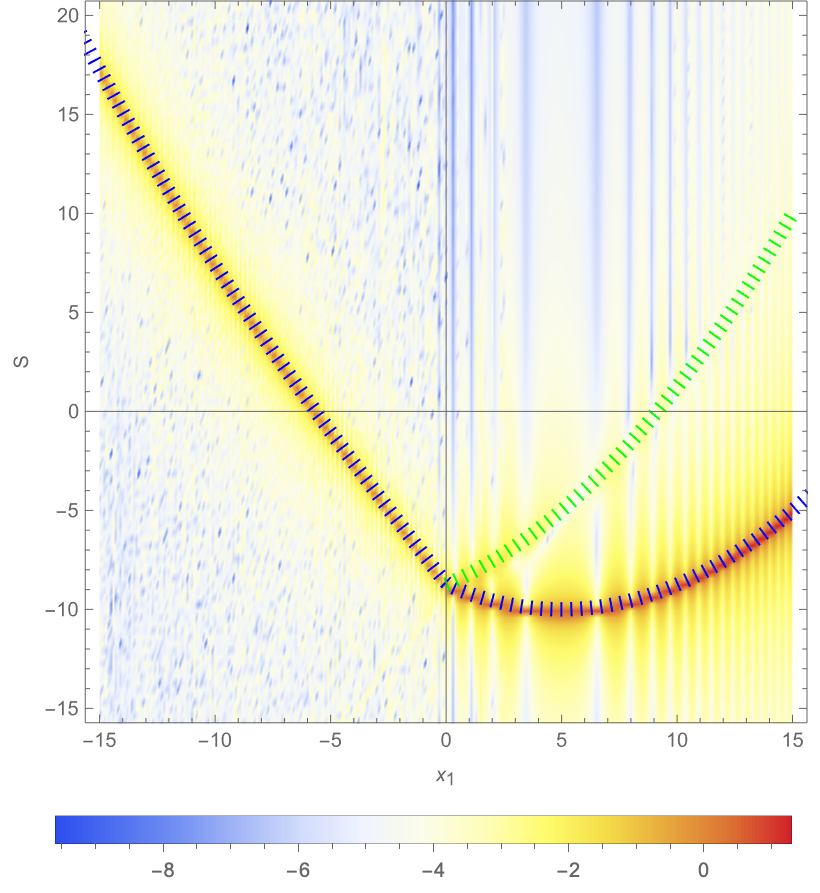}
        \caption{$x_0=5$}\label{fig:fourier_smooth_R}
    \end{subfigure}
    \caption{A logarithmic plot of the Fourier transform of the smooth step propagator with respect to the frequency parameter $\omega = 1/ \hbar$ for the step height $V_0=1$, steepness $\alpha=5$, mass $m=1$ and propagation time $T=10$. The blue dashed lines correspond to the actions of the real classical paths. The dashed vertical lines mark the caustics.}\label{fig:fourier_smooth}
\end{figure*}

When evaluating the spectrum for a fixed initial position as a function of the final position, we can track the classical paths relevant to the path integral of the smooth and Heaviside step (see figs.\ \ref{fig:fourier} and \ref{fig:fourier_smooth}). The blue curves correspond to the actions of the real classical paths. For $x_0=-5$, we observe the presence of two caustics bounding the region with three real classical paths. Out of the left fold caustics, we observe the emergence of a complex semi-classical path represented by the green curve. The peak in the spectrum corresponding to the complex semi-classical paths is constant in space, reflecting the observation that the associated classical action is real. For the Heaviside step, the quantum reflections persist in the semi-classical limit.

\bigskip
Next, consider the smooth step. Specifically, we consider the transition amplitude for a particle of mass $m=1$ to propagate from $x_0= -5$, to $x_1 = -9.25$ in a smooth step potential with the strength $V_0=1$ and steepness parameter $\alpha=5$. The system has three interesting candidate solutions to the boundary value problem. The first path is a real classical solution bouncing of the step with the action $S[x_C]\approx 0.909$ (the real saddle). In addition, we identify a complex solution which is the analytic continuation of the complex semi-classical path emerging at the fold caustic with the classical action $S[x_C]\approx 10.384 + 0.256 i$ (the caustic saddle). Finally, we identify the analytic continuation of the solution suggested by the reflection rates with the classical action $S[x_C]\approx 10.555$ (the topological saddle). Note that the action of the topological saddle point is real. 

The spectrum of the propagator shows two peaks in agreement with the real saddle point and either the caustic or the topological saddle point. From theoretical grounds, the caustic saddle point is relevant, as we have not crossed a Stokes line. However, it is difficult to infer the relevance of the topological saddle point from the Fourier analysis, as their real parts are very similar.

When evaluating the spectrum as a function of the final position $x_1$ we obtain a pattern resembling the spectrum of the Heaviside step (see fig.\ \ref{fig:fourier_smooth}). The signal corresponding to the semi-classical path decays as we decrease $x_1$, signalling that the imaginary part of the action grows. The transition from the left to the right of the step is only performed in a classical fashion. There is no quantum tunnelling at play. When both $x_0$ and $x_1$ are on the right side of the step, we observe the presence of two classical paths. The lower arc represents the real propagation, resembling a free particle. The upper arc corresponds to a quantum reflection.

\subsection{The Laplace transform}
In the previous section, we considered three candidate solutions to the boundary value problem propagating in the smooth step on the left of the step. We were unable to infer the relevance of the topological saddle point from the Fourier spectrum, as the real parts of the classical action of the caustic and topological saddle point are very similar. We lift this degeneracy with the Laplace transform of the propagator, inspired by the early work \cite{Balian:1974}.

The Laplace transform of the propagator with respect to $\omega$ yields
\begin{align}
    \mathcal{L}(s) &= \int_A^B \sqrt{\frac{2 \pi}{i \omega}}G(x_1,x_0;T) e^{-\omega s}\mathrm{d}\omega\\
    &\approx \Theta(T)
    \sum_{x_C } \sqrt{\frac{\partial^2 S_C}{\partial x_0 \partial x_1}} \int_A^B e^{i \omega S_C} e^{- \omega s}\mathrm{d}\omega\\
    &= \Theta(T)\sum_{x_C } \sqrt{\frac{\partial^2 S_C}{\partial x_0 \partial x_1}} \frac{e^{A(i S_C - s)} - e^{B(i S_C - s)}}{s - i S_C}
\end{align}
Since
\begin{align}
    \left| \frac{1}{s- i S_C }\right|^2 = \frac{1}{(\text{Im}[S_C]+ s)^2 + \text{Re}[S_C]^2}
\end{align}
the modulus of the Laplace transform $|\mathcal{L}(s)|^2$ can be used to infer the imaginary parts of the classical actions of the relevant (semi-)classical paths. At first sight, one might expect the Laplace transform to peak at the imaginary parts of the actions. However, as $s$ is positive and the Picard-Lefschetz theory demonstrates that the imaginary parts of the relevant semi-classical paths are always positive, the Laplace transform $|\mathcal{L}(s)|^2$ typically peaks for some small positive $s$ due to the real classical paths and decays for large $s$.

\begin{figure}
    \centering
    \includegraphics[width=\linewidth]{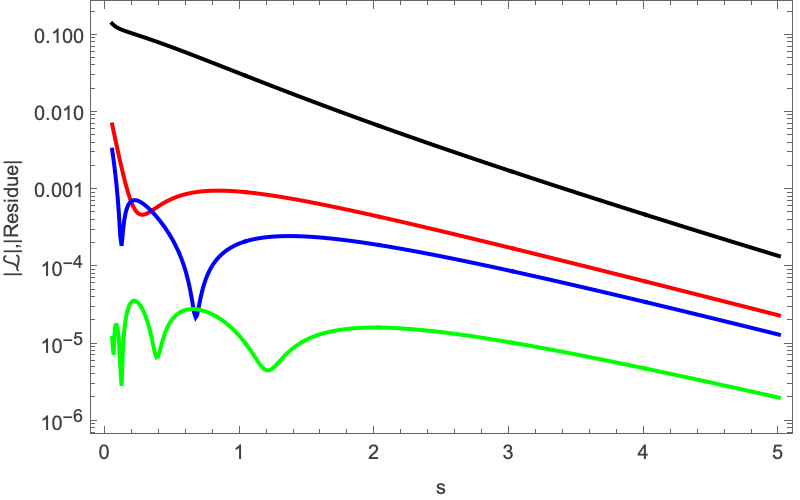}
    \caption{The residue of the Laplace transform modelled by the WKB approximation, including only the real classical path (red), the real classical path and the semi-classical path associated with the caustic (blue), and the real classical path, the semi-classical path associated with the caustic and the topological classical path (green). We compare the residue with the modulus of the Laplace transform $|\mathcal{L}(s)|$ (black). In this calculation, we use the configuration $x_0 = -5$, $x_1 = -9.25$, with a propagation time $T = 10$ for the smooth step with the strength of the step $V_0=1$ and the steepness parameter $\alpha = 5$. The real, caustic saddle and topological saddles lead to the classical actions $S[x_C] \approx 0.909$, $S[x_C]\approx 10.384 + 0.256 i$ and $S[x_C]\approx 10.555$.}\label{fig:residue}
\end{figure}

To demonstrate this method, we consider the smooth step, with boundary conditions $x_0$ and $x_1$ on the left of the potential discussed in the previous section. We find that in addition to the real classical path and the semi-classical path associated with the caustic, also the topological semi-classical path associated with the quantum reflection rate is relevant to the path integral (see fig.\ \ref{fig:residue}). Remarkably, the topological semi-classical path has a real action at $E=V_0$ as well as $\alpha \rightarrow \infty$. Its contribution to the propagator is unsuppressed as we approach the semi-classical limit $\hbar \to 0$. The real-time transition amplitude of physically reasonable theories can seemingly include contributions of complex-semiclassical paths with real actions that are not exponentially suppressed for small Planck constants, beyond the pathological Heaviside step potential.  

\subsection{Experimental remarks}
In experiments, it is not possible to change the Planck constant. However, using the fact that the path integral is preserved under the scaling $x_0 \mapsto x_0/C, x_1 \mapsto x_1/C, \alpha \mapsto C\alpha, \hbar \mapsto \hbar /C, T \mapsto T/C$, for positive $C$, we can effectively vary the Planck constant by changing the boundary conditions and propagation time. Concretely, the propagator 
\begin{align}
    G(x_1,x_0;T)
\end{align}
with steepness $\alpha$ at Planck constant $\hbar = C$ is equivalence to 
\begin{align}
    G(x_1/C, x_0/C; T/C)
\end{align}
with steepness $C \alpha$ at Planck constant $\hbar = 1$. In practice, we propose to measure the phase of the propagator with holography by interfering an unperturbed particle with a particle encountering the step.

%%%%%%%%%%%%%%%%%%%%%%%%%%%%%%%
\section{Applications}\label{sec:applications}
In the present paper, we restrict ourselves to the study of non-relativistic particles in 
simple step potentials. This setting is an ideal sandbox to study the real-time path integrals, as the theory is more generic than the harmonic oscillator, yet both the propagator and classical paths can be solved in closed (admittedly implicit) form. We believe this work can have an impact in many fields in physics and mathematics. In the next sections, we will briefly discuss a couple of applications of our work to different fields in physics and mathematics.

\subsection{Anti-reflective Coating}
Anti-reflective or antiglare coatings are a type of optical coating applied to the surface of lenses, other optical elements, and photovoltaic cells to reduce reflection. The coating improves the efficiency of lenses as less light is lost due to reflection. In complex systems such as cameras, binoculars, telescopes, and microscopes, the reduction in reflections also improves the contrast of the image. The first anti-reflective coating was discovered by Lord Rayleigh in 1886. The optical glass available at the time develops a tarnish over time, due to chemical reactions with air. Rayleigh tested some old, slightly tarnished pieces of glass and found to his surprise that they transmitted more light than new, clean pieces. In modern coatings, layer thicknesses are selected to induce destructive interference in the reflected beam and constructive interference in the transmitted beams \cite{Raut:2011, Buskens:2016}.

The development of anti-reflective coatings is an active field of research. The present study of quantum reflections in step potentials, though complex semi-classical paths, may improve our understanding of these coatings and ultimately improve the properties of next-generation coatings.

\subsection{Laser Physics}
In attosecond laser physics, it is common to study the ionisation and recombination of electrons. Light-matter interactions in strong laser physics, such as high-harmonic generation (HHG), are naturally described in terms of highly-oscillatory conditionally convergent integrals over the ionisation and recombination times \cite{Pisanty:2020, Weber:2025}. Traditionally, these highly oscillatory integrals have been approximated with saddle-point methods. However, the saddle points of interest are always complex and determining the relevancy of the saddle point has remained an open problem.  Numerical experiments with complex orbits resulted in quantitative success \cite{Balian:1978, salieres:2001, Milosevic:2024}.  The prescriptions had been heuristic, sometimes using complex time, and at other times complex positions, without a unified description of the equivalence class described in this paper.

On the one hand, Picard-Lefschetz theory, the caustic curves and Stokes lines will likely provide a theoretical solution to this problem \cite{Weber:2025b}. On the other hand, the Fourier and Laplace methods proposed in this work may yield an experimental way to identify relevant saddle points. By measuring the intensity of emitted radiation while changing the external parameters of the experiment, we may effectively vary the Planck constant. The Fourier transform will peak at the real part of the action of the relevant saddle points. The Laplace transform may further highlight the relevant saddle points through the imaginary part of the actions in the saddle points.

\subsection{Particle Creation}
Quantum fields on curved spacetime can famously lead to the creation of particles. To study this phenomenon, Birrell and Davis consider a scalar field in a homogeneous and isotropic Friedmann-Lemaître-Robertson-Walker (FLRW) universe (see section 3.4 of \cite{Birrell:1984}). Assuming that the square metric $a(\eta)^2$ evolves as a smooth step, 
\begin{align}
    a(\eta)^2 = A + B \tanh (\rho \eta)
\end{align}
in conformal time $\eta$ for constant $A, B, \rho$, they derive the eigenstates (\ref{eqn:tanheig}) and the reflection analogue of particle creation rates. The equation of the modes in quantum field theory (see equation (3.86) in \cite{Birrell:1984}) is equivalent to the time-independent Schr\"odinger equation of a non-relativistic particle in a Woods-Saxon potential with the spatial coordinate $x$ replaced by conformal time $\eta$, $\alpha \rightarrow \rho$, mass rescaled to $m\rightarrow m\sqrt{2}$, height $V_0\rightarrow -B/2$, $\hbar=1$ and offset $A=B$. They use the reversed convention, that the in state is on the top of the potential, while the out state is down the step.  In quantum field theory, each in-state eigenmode is in the vacuum corresponding to (\ref{eqn:in}), and the Bogolubov coefficients are the projections onto the out states (\ref{eqn:out}). Their particle creation rates (see equations (3.94) and (3.95) in \cite{Birrell:1984}) are equivalent to the reflection and tunnelling rates obtained for the smooth step \eqref{eqn:RT}.
    
The creation of particles is exponentially suppressed. Historically, this was described as a non-adiabatic process creating particles from the vacuum.  While the apparent exponential dependence of the creation process begs for an instanton process, to our knowledge, none have been identified \footnote{Prokopec, private communication}.  The new saddles provide an explicit construction for the creation of particles in cosmological expanding spacetimes. The same instanton gives a physical picture of the process.

\bigskip
Alternatively, particle creation in curved spacetime can be studied from the worldline perspective. Hawking and Hartle \cite{Hartle:1976} famously studied black-hole radiance using the path integral over worldlines and provided an alternative derivation for Hawking radiation and the Hawking temperature. In their derivation, they studied the path integral of a particle in a Schwarzschild spacetime and used both Wick rotations to imaginary time and analytic continuations of paths past the curvature singularity at $r=0$. The particle creation rate emerges through the exponential suppression of the emission rate with respect to the absorption rate. This suppression is again likely to emerge from a complex saddle point. 

The techniques developed in this paper encourage a deeper study of the path integral over worldlines in black hole spacetimes. Whereas the step potential is only singular when analytically continued into the complex plane, the curvature singularity of the Schwarzschild black hole spacetime exists for strictly real configurations. It seems likely that a rigorous instanton explanation of Hawking radiation in real time will require the equivalence classes of complex classical paths proposed for the step potential, as these will allow us to continue the action beyond the singularity crossings. The Fourier and Laplace methods may very well prove useful to infer the relevance of the instantons.

\subsection{Quantum Cosmology and Gravity}
The path integral for gravity and its application to quantum gravity and quantum cosmology is often studied in terms of complex saddle points, instantons or complex semi-classical paths \cite{Hartle:1983, Vilenkin:1986,  Lehners:2023}. This path integral generally consists of a term resembling a path integral from quantum mechanics and an additional integral over the lapse \cite{Teitelboim:1983}. The relevance of the saddle points is often a topic of debate. 

The present study suggests that when the equation of motion is not of Painlevé type, we need to extend the list of candidate instantons to include equivalence classes corresponding to the analytic continuation of complex semi-classical paths. The relevance of these instantons can be explored using the caustic and Stokes lines discussed in this paper. When it is possible to numerically evaluate the path integral for mini-superspace models, the Fourier and Laplace transform techniques can provide further evidence of the relevance of these instantons.

\subsection{Resurgence}
In the present study, we approximated the path integral with the WKB approximation. However, this is but the first term in a small $\hbar$ expansion of the path integral. In this expansion, known as a trans-series, each relevant saddle point leads to an exponential term of the form $e^{i S/\hbar}$ and a series expansion in powers of $\hbar$. Unfortunately, these series expansions are typically only asymptotic. Their first approach to the propagator is to a point after which the series diverges. The mathematical properties of these trans-series, their Borel resummation and approximations through hyperasymptotic expansions approximations is studied in the mathematical theory of resurgence \cite{Dingle:1973, Berry:1991, Aniceto:2019, Dorigoni:2019}. 

The techniques of resurgence have, over the last years, been increasingly often applied to problems in quantum physics, as they provide a way to study the asymptotic series emerging in perturbation theory. Present studies of resurgence in quantum physics have largely restricted attention to solutions of the time-independent Schrödinger equation (see, for example \cite{Dunne:2014}). If we are to develop the trans-series expansion of real-time quantum physics through the Feynman path integral for general potentials, the present analysis suggests that we need to include the equivalence classes of complex saddle points proposed in this paper. By resumming the associated series expansions, we may penetrate the structure of the real-time path integral in an analytic fashion. In an upcoming paper, we evaluate the trans-series expansion of a non-relativistic particle in a Woods-Saxon potential  \cite{Feldbrugge:Ines}.

%%%%%%%%%%%%%%%%%%%%%%%%%%%%%%%
\section{Conclusion}\label{sec:conclusion}
We derive the spectral representation of the real-time Feynman path integral of a non-relativistic particle evolving in a Woods-Saxon and a Heaviside step potential. We, for the first time, evaluate the propagator and identify the relevant real and complex (semi-)classical paths. To explain the behaviour of the propagator in detail, we identify the caustics and Stokes lines in configuration space and extend the notion of semi-classical paths or instantons to equivalence classes of classical paths, where not only the path but also the time domain is smoothly deformed into complex space while avoiding the crossings of singularities. We conclude that, whereas in classical theory the paths are fundamental, in quantum theory only the classical actions count, as we identify configurations for which a contribution to the propagator cannot be explained using a complex-classical path over the real interval $[0,T]$.

The identification of relevant instantons or semi-classical paths is an open problem in many fields of physics. In theory, Picard-Lefschetz theory promises to identify the relevant complex semi-classical paths with a simple intersection criterion. However, this criterion is hard to implement in practice for all but a few physical theories. The present paper suggests that the problem of identifying the relevant instantons is more tricky than expected; Besides determining the relevancy of known instantons, we need to include analytic continuations of instantons that were previously missed. To shed light on the problem and develop a deeper understanding, we here propose to infer the relevant semi-classical from the path integral itself (rather than using the classical paths to approximate it). We first evaluate the propagator numerically, using either the spectral representation or novel numerical methods \cite{Feldbrugge:2025}, and subsequently Fourier and Laplace transform to propagator with respect to the reciprocal of the Planck constant $\omega = 1/\hbar$. Relevant classical paths are identified as peaks in the Fourier spectrum centred at the real parts of the associated classical actions. The Laplace transform enables the estimation of the relevant classical paths using the imaginary parts of the classical actions.

Complex semi-classical actions are normally assumed to yield exponentially small contributions in reasonable experimental settings (corresponding to small Planck constants). Yet in both the smooth Woods-Saxon and discontinuous Heaviside step potential, we have found that relevant complex semi-classical paths may have real classical actions. This suggests that their contribution will persist in the semi-classical limit and enriches the field of non-perturbative quantum physics. 

Though the present paper restricts the analysis to step potentials, we believe that these realisations generalise and can have an impact in many fields in physics and mathematics. In the future, we hope to use the present work to study the importance of analytically continued classical paths in fields ranging from laser physics, quantum gravity, to the mathematical field of asymptotic series and resurgence.

%%%%%%%%%%%%%%%%%%%%%%%%%%%%%%%
\section*{Acknowledgments}
The work of JF is supported by the STFC Consolidated Grant ‘Particle Physics at the Higgs Centre,’ and, respectively, by a Higgs Fellowship and the Higgs Chair of Theoretical Physics at the University of Edinburgh. 

ULP thanks Tomislav Prokopec for helpful insights on the curved space particle creation.

For the purpose of open access, the authors have applied a Creative Commons Attribution (CC BY) license to any Author Accepted Manuscript version arising from this submission.

\bibliographystyle{apsrev4-1}
\bibliography{library.bib}

\appendix

%%%%%%%%%%%%%%%%%%%%%%%
% \section{Appendix}

\end{document}